\RequirePackage[colorlinks,allcolors=blue]{hyperref} 

\documentclass[draft]{agujournal2019}
\usepackage{url} 
\usepackage{lineno}
\usepackage[inline]{trackchanges} 

\usepackage{soul}
\usepackage{graphicx}
\usepackage{textcomp, gensymb}
\usepackage{caption}
\usepackage{subcaption}
\usepackage[export]{adjustbox}
\usepackage{siunitx}

\draftfalse

\journalname{JGR Planets}

\begin{document}


\title{The Thermal Structure and Composition of Jupiter's Great Red Spot from JWST/MIRI}
%

\authors{
Jake Harkett\affil{1},
Leigh N. Fletcher\affil{1},
Oliver R.T. King\affil{1},
Michael T. Roman\affil{1},
Henrik Melin\affil{1},
Heidi B. Hammel\affil{2},
Ricardo Hueso\affil{3},
Agustín Sánchez-Lavega\affil{3},
Michael H. Wong\affil{4},
Stefanie N. Milam\affil{5},
Glenn S. Orton\affil{6},
Katherine de Kleer\affil{7},
Patrick G.J. Irwin\affil{8},
Imke de Pater\affil{9},
Thierry Fouchet\affil{10},
Pablo Rodríguez-Ovalle\affil{10},
Patrick M. Fry\affil{11},
Mark R. Showalter\affil{12}
}

\affiliation{1}{School of Physics and Astronomy, University of Leicester, University Road, Leicester LE1 7RH, UK}
\affiliation{2}{Association of Universities for Research in Astronomy, Suite 1475, 1331 Pennsylvania Avenue NW, Washington DC 20004, USA}
\affiliation{3}{Escuela de Ingeniería de Bilbao, Universidad del País Vasco, UPV/EHU, Bilbao, Spain}
\affiliation{4}{Center for Integrative Planetary Science, University of California, Berkeley, CA
94720, USA}
\affiliation{5}{Astrochemistry Laboratory Code 691, NASA Goddard Space Flight Center, 8800 Greenbelt Road, Greenbelt MD 20771, USA}
\affiliation{6}{Jet Propulsion Laboratory, California Institute of Technology, Pasadena, CA 91109, USA}
\affiliation{7}{Division of Geological and Planetary Sciences, California Institute of Technology, Pasadena, CA 91125, USA}
\affiliation{8}{Department of Physics, University of Oxford, Parks Rd, Oxford, OX1 3PU, UK}
\affiliation{9}{Department of Astronomy, University of California, Berkeley, California 94720, USA}
\affiliation{10}{LESIA, Observatoire de Paris, Université PSL, CNRS, Sorbonne Université, Université Paris-Cité, Meudon, France}
\affiliation{11}{Space Science and Engineering Center, University of Wisconsin, 1225 west Dayton Street, Madison, WI, 53706, USA}
\affiliation{12}{SETI Institute, 189 Bernardo Ave., Mountain View, CA 94043, USA}



\correspondingauthor{Jake Harkett}{jh852@leicester.ac.uk}




\begin{keypoints}
\item Jupiter’s Great Red Spot was observed by JWST/MIRI in 2022 to study the 3D structure of temperature, aerosols and gaseous species.
\item A series of stratospheric hot-spots were discovered to be co-moving with the Great Red Spot in the longitude direction.
\item Elevated phosphine and a complex distribution of ammonia gas were observed inside the Great Red Spot.
\end{keypoints}

%
%


\begin{abstract}
    Jupiter’s Great Red Spot (GRS) was mapped by the JWST/Mid-Infrared Instrument (4.9 - 27.9 µm) in July and August 2022. These observations took place alongside a suite of visual and infrared observations from; Hubble, JWST/NIRCam, VLT/VISIR and amateur observers which provided both spatial and temporal context across the jovian disc. The stratospheric temperature structure retrieved using the NEMESIS software revealed a series of hot-spots above the GRS. These could be the consequence of GRS-induced wave activity. In the troposphere, the temperature structure was used to derive the thermal wind structure of the GRS vortex. These winds were only consistent with the independently determined wind field by JWST/NIRCam at 240 mbar if the altitude of the Hubble-derived winds were located around 1,200 mbar, considerably deeper than previously assumed. No enhancement in ammonia was found within the GRS but a link between elevated aerosol and phosphine abundances was observed within this region. North-south asymmetries were observed in the retrieved temperature, ammonia, phosphine and aerosol structure, consistent with the GRS tilting in the north-south direction. Finally, a small storm was captured north-west of the GRS that displayed a considerable excess in retrieved phosphine abundance, suggestive of vigorous convection. Despite this, no ammonia ice was detected in this region. The novelty of JWST required us to develop custom-made software to resolve challenges in calibration of the data. This involved the derivation of the `FLT-5' wavelength calibration solution that has subsequently been integrated into the standard calibration pipeline.
\end{abstract}


\section*{Plain Language Summary}
    Regularly observed for over 150 years, Jupiter’s Great Red Spot (GRS) is one of the best documented storms in the Solar System. Despite the frequency of observations, crucial questions remain unanswered regarding the internal composition, dynamics and driving mechanism for the storm. Mid-infrared observations acquired by the James Webb Space Telescope allowed us to peer past the colourful clouds to assess the composition and dynamics of the jovian weather layer. Data from the mid-infrared instrument was modelled in the 7.30 – 10.75 µm range to `retrieve’ the temperature distribution. Further modelling of these temperatures allowed us to derive the wind speeds throughout the vortex, enabling us to assess the dynamics of the GRS and how the vortex interacts with its surroundings. In addition, a series of hot-spots were observed above the GRS that could be the result of atmospheric wave activity. The distribution of ammonia was also mapped in this spectral range. Comparison of the distribution of ammonia to the cloud distribution implied that this molecule may condense into the thick layers of cloud above the GRS. Finally phosphine, indicative of upwelling air was mapped and allowed us to identify a small convective storm north-west of the GRS.

%
%

\section{Introduction}
    
    The Great Red Spot (GRS) is a remarkable feature of the jovian atmosphere. Being both the largest and the longest-lived anticyclone in the Solar System, it dominates the dynamics of the entire southern hemisphere of Jupiter from its location within the Southern Tropical Zone (STrZ). Yet there are still numerous unanswered questions regarding the dynamical and compositional structure as well as the formation, evolution and potential vertical decay processes. By understanding this most extreme of meteorological features, we can begin to unravel many of the unknowns regarding other giant planet vortices. We have utilized James Webb Space Telescope (JWST) mid-infrared spectroscopy from the Mid-InfraRed Instrument (MIRI) to produce 3D maps of both the thermal and compositional structure within the GRS. These observations occurred in July and August 2022, at the very start of the JWST science phase. This allowed us to test the capabilities of this new observatory to create local maps of highly dynamic features on a bright, rotating and non-sidereal extended source. The JWST/MIRI observations were taken on similar dates to visual Hubble/WFC3 and near-infrared JWST/NIRCam imaging in order to correlate any deeper compositional or dynamical changes in the vortex with disturbances to the aerosol population visualised in reflected sunlight. These supporting observations also provided crucial spatial context of the atmosphere outside the GRS, allowing us to monitor the effect of the GRS dynamics on the surroundings. A timing issue with the execution of the MIRI observations resulted in us receiving two sets of mid-infrared data of the GRS separated by two weeks, allowing us to determine which features within the vortex are stable on timescales longer than the GRS rotation time of approximately 4 days. These investigations were motivated by (a) a need for an understanding of the vertical thermal and compositional structure of the GRS, something that can only be constrained through broad wavelength high-resolution spectroscopy and (b) a lack of knowledge regarding the coupling between the internal dynamics of the GRS and the surrounding atmosphere, particularly in the stratosphere above.

    \subsection{Visual and Near-infrared observations of the Great Red Spot}
        Fig. \ref{jup_rgb_troposphere}a and b display the visual context data acquired for this study, with the positions of major atmospheric features labelled. Visual imaging of the GRS reveals the distinctive red colouration of the vortex, produced by an unidentified red chromophore assumed, but not confirmed to be the same molecule responsible for the red colouration of Jupiter's belts \cite{baines_2019grschromophore_paper}. Near-infrared imaging of the vortex at 890 nm \cite{sanchez-lavega_2018grsjunocam_paper, baines_2019grschromophore_paper, sanchez-lavega_2021grsinteractions_paper} displays a region of high aerosol reflectance within the GRS. Since this spectral region is dominated by CH$_4$ absorption at $\sim$100 mbar \cite{dahl_2021grscolour_paper}, such observations suggest that the aerosol population within the GRS is elevated to the upper troposphere or even the lower stratosphere. Knowledge of the winds within the vortex have benefited from a series of annual 631 nm observations made using Hubble as part of the Outer Planet Atmosphere Legacy (OPAL) \cite{simon_opal_paper} and Wide Field Coverage for Juno (WFCJ) programme \cite{wong_wfcj_paper}. Approximately 14,000 $\times$ 11,000 km in size (2018 measurement) \cite{sanchez-lavega_2021grsinteractions_paper}, the majority of the velocity and vorticity of the system is located in a high-velocity anticyclonic ring originating at 70-85\% of the radius of the GRS \cite{sada_1996grsvoyager_paper, simon-miller_2002grs_paper, choi_2007grsvorticity_paper, wong_2021grsvelocities_paper}. This peripheral jet encapsulates the cold, anticyclonic temperature anomaly and prevents mixing of this air with the surroundings \cite{fletcher_2010grs_paper}. The vorticity forces the two jets bounding the STrZ (flowing retrograde at 19.5$\degree$S and prograde at 26.5$\degree$S planetographic latitude respectively) to flow around the vortex, separating the red chromophore of the GRS from the surrounding belts \cite{7}. These jets collide north-west of the GRS, generating a turbulent region of vigorous convective activity known as the GRS wake. Patchy white clouds in this region have been identified as thunderstorms due to the detection of lightning by Galileo \cite{gierasch_2000galileolightning_paper}. Surrounding the GRS is a high-albedo ring known as the hollow \cite{sanchez-lavega_2018grsjunocam_paper}. This is further enclosed by a dark, cloud free periphery of low opacity \cite{de-pater_2010vortexrings_paper}. The anticyclonic circulation of the GRS and interactions with the jovian zonal wind system can cause material from the hollow to be swept out into a triangular shape north of the GRS with an equatorial vertex at $\sim$15$\degree$S. \cite{rogers_1995jupiter_book}. It is possible for the dark line around the hollow to break at this point and release a tail of white clouds into the surrounding Southern Equatorial Belt (SEB) that has been named a `chimney' \cite{anguiano-arteaga_2021grsaerosol_paper, anguiano-arteaga_2023grsaerosol_paper}.

    \subsection{Mid-infrared observations of the Great Red Spot}
        The early history of mid-infrared observations for each of the Giant Planets is reviewed in \citeA{roman_giantplanets_review}. Both Voyagers 1 and 2 flew by Jupiter in 1979 carrying the InfraRed Interferometer Spectrometer (IRIS) instrument, containing a Michelson interferometer operating in the 2.5 - 55 µm spectral range with a spectral resolution of R = 42 - 558 \cite{ww}. These observations confirmed the GRS was a cold-core anticyclonic feature surrounded by a warm, cyclonic annulus \cite{xx, flasar_1981grsvoyager_paper}. Thermal wind analysis saw the anticyclonic winds decay with altitude into the lower stratosphere \cite{flasar_1981grsvoyager_paper}. Cloud clearing was observed within the cyclonic annulus at 5 µm suggesting that localised upwelling within the GRS causes subsidence in this periphery. This was further supported by the observation of a localised para-H$_2$ minimum within the GRS vortex, indicating vigorous upwelling occurring at a rate faster than the ortho-para H$_2$ ratio could reset to thermal equilibrium \cite{sada_1996grsvoyager_paper, simon-miller_2002grs_paper}. The distributions of NH$_3$ and PH$_3$ were also mapped for the first time. The retrieved abundance of these gases was ambiguous though, with IRIS detecting a localised depletion in NH$_3$ and no difference in abundance compared to the surroundings for PH$_3$ \cite{bb}.  Finally a localised maxima in aerosol opacity was detected within the GRS at both 5 µm and 44 µm \cite{sada_1996grsvoyager_paper}.
        \par
        The atmosphere of Jupiter was further analysed over 35 orbits by the Galileo orbiter (1995 - 2003). The PhotoPolarimeter-Radiometer (PPR) instrument \cite{zz} observed the GRS in five discrete mid/far-infrared bands between 15-100 µm. This allowed the 700-200 mbar temperature to be mapped, indicating that the cold temperature anomaly of the GRS vortex persists at a range of altitudes, typically being 3 K lower than the regions east and west of the vortex \cite{qq, fletcher_2010grs_paper}. Such a broad low temperature anomaly implies de-spinning of the vortex with altitude according to the thermal wind equations \cite{flasar_1981grsvoyager_paper}. These velocities are predicted to increase with depth to a maximum at the mid-plane \cite{de-pater_2010vortexrings_paper, palotai_2014anticyclonemodelling_paper, 60}, where the low-temperature anomaly is expected to become indistinguishable from the surroundings. This mid-plane has never been detected. There was also tentative evidence of a north-south temperature asymmetry within the GRS and a warm periphery south of the GRS, corresponding to a dark region at visible wavelengths \cite{fletcher_2010grs_paper}. However, the low spatial resolution of the instrument made these observations difficult to confirm. In addition to the PPR, Galileo also carried the Near-Infrared Mapping Spectrometer (NIMS) \cite{jjj}. This observed the GRS and surrounding atmosphere in the 0.7 - 5.2 µm spectral range, allowing observations of the 5.0 µm `spectral window', where a dearth of CH$_4$ and H$_2$ opacities allows observation of deep tropospheric thermal emission (5-10 bar) with higher-altitude clouds silhouetted against the brighter thermal background. This confirmed that the brightest 5 µm regions of Jupiter's atmosphere corresponded to the darkest regions in the visible spectrum. NH$_3$ ice clouds were also spectroscopically identified for the first time in the turbulent GRS wake through detection of the 2.96 µm $\nu_3$ NH$_3$ ice absorption band \cite{baines_nh3icev3_paper}.
        \par
        In 2000, Cassini-Huygens observed Jupiter over a period of 6 months while en-route to Saturn. Cassini carried the Composite InfraRed Spectrometer (CIRS) instrument \cite{ccc}, obtaining continuous spectra in the 7.1 - 1,000 µm range with a spectral resolution of R = 20 - 2,800. This enabled detection of localised enhancements in both PH$_3$ and NH$_3$ within the GRS \cite{irwin_2004ph3cirs_paper, ddd, fletcher_2010grs_paper}, disagreeing with previous spacecraft observations of these species. The spatial resolution of CIRS was also sufficient to detect a north-south asymmetry of NH$_3$ in the GRS periphery, with a localised enhancement in the north coinciding with an arc of lower temperatures. This along with reassessments of the Voyager and Galileo data \cite{sada_1996grsvoyager_paper, eee, m, simon-miller_2002grs_paper, fff} suggested that the GRS tilts in the north-south direction, with stronger upwelling in the northern periphery than the south. CIRS also detected further spectral signatures of the NH$_3$ cloud layer in the GRS wake, this time through detection of the 9.46 µm $\nu_2$ NH$_3$ ice absorption band \cite{wong_nh3icev2_paper}. This is difficult to observe from the ground due to terrestrial O$_3$ absorption at 10.0 µm. 
        \par
        These visitations by spacecraft have offered glimpses into the internal structure of the GRS. However, the nature of their trajectory either as a flyby or a close-orbiting probe has meant that these observations are limited to brief snapshots separated by years. The increasing size and complexity of ground-based observatories has meant that the morphology of the GRS can now be continuously studied on longer timescales. Such large telescopes can often exceed the spatial resolution of past instruments such as CIRS. The spectral resolution of these instruments can also surpass 10,000, allowing the spectra of the GRS to be probed in unprecedented detail. However, these high-resolution spectra are usually obtained in a series of narrow filters (9 in the case of NASA’s 3 m InfraRed Telescope Facility (IRTF/TEXES) \cite{lacy_2002texes_paper}). Furthermore, telluric absorption results in the 5 - 11 µm spectral range being difficult to calibrate and the 5.5 - 7.7 µm range being impossible to observe from the ground. Although IRTF/TEXES lacks the spatial resolution to observe spatial variation within the GRS vortex, it concurred with the Cassini/CIRS observation of a north-south asymmetry in the NH$_3$ distribution and that PH$_3$ is in excess over the GRS \cite{fletcher_2016texes_paper}. These observations also confirmed the compositional anomalies were confined to the troposphere and that they also appeared to be becoming more circular over time, consistent with the observed shrinkage of the GRS \cite{7}. The greater spatial resolution of the 8.2 m Very Large Telescope (VLT) Imager and Spectrometer for mid-InfraRed (VLT/VISIR) \cite{lagage_2004visir_paper} enabled the observation of the weakly-cyclonic warm core within the vortex of the GRS \cite{fletcher_2010grs_paper}.
        
    \subsection{JWST observations of the Great Red Spot}
        Although the previous Mid-infrared observations of the GRS have enabled the broad thermal and compositional structure to be revealed, the retrieved gaseous distributions across the GRS have been ambiguous due to different studies producing different results. Also, the majority of these observations are limited to a handful of discrete wavelength bands, translating into discrete altitudes. Cassini/CIRS did provide broader wavelength coverage, but the spatial resolution was poorer. With even broader wavelength coverage of 4.9 - 27.9 µm and a resolving power ranging from R = 3,710 - 1,330, JWST offers a clear advance over any previous facilities and will provide the most comprehensive mapping available of the vertical temperature structure, the windshear, stability, gaseous and aerosol distributions. In addition to this, JWST offers a window to connect the cold temperature anomaly of the tropospheric anticyclone to the dynamics of the overlying stratosphere. The 3D molecular maps in particular have the potential to resolve many of the ambiguities observed in corresponding 2D maps produced in the past.
        \par
        The structure of the paper is as follows: Section \ref{data_methods} describes the data and the post-processing steps applied to the MIRI observations. Section \ref{NEMESIS} outlines our spectral modelling approach. Section \ref{results} presents the results from this modelling. Section \ref{discussion} analyses these results and derives the resulting thermal winds and Section \ref{conclusions} concludes the study.

    \begin{figure}
        \centering
        \includegraphics[width=\textwidth]{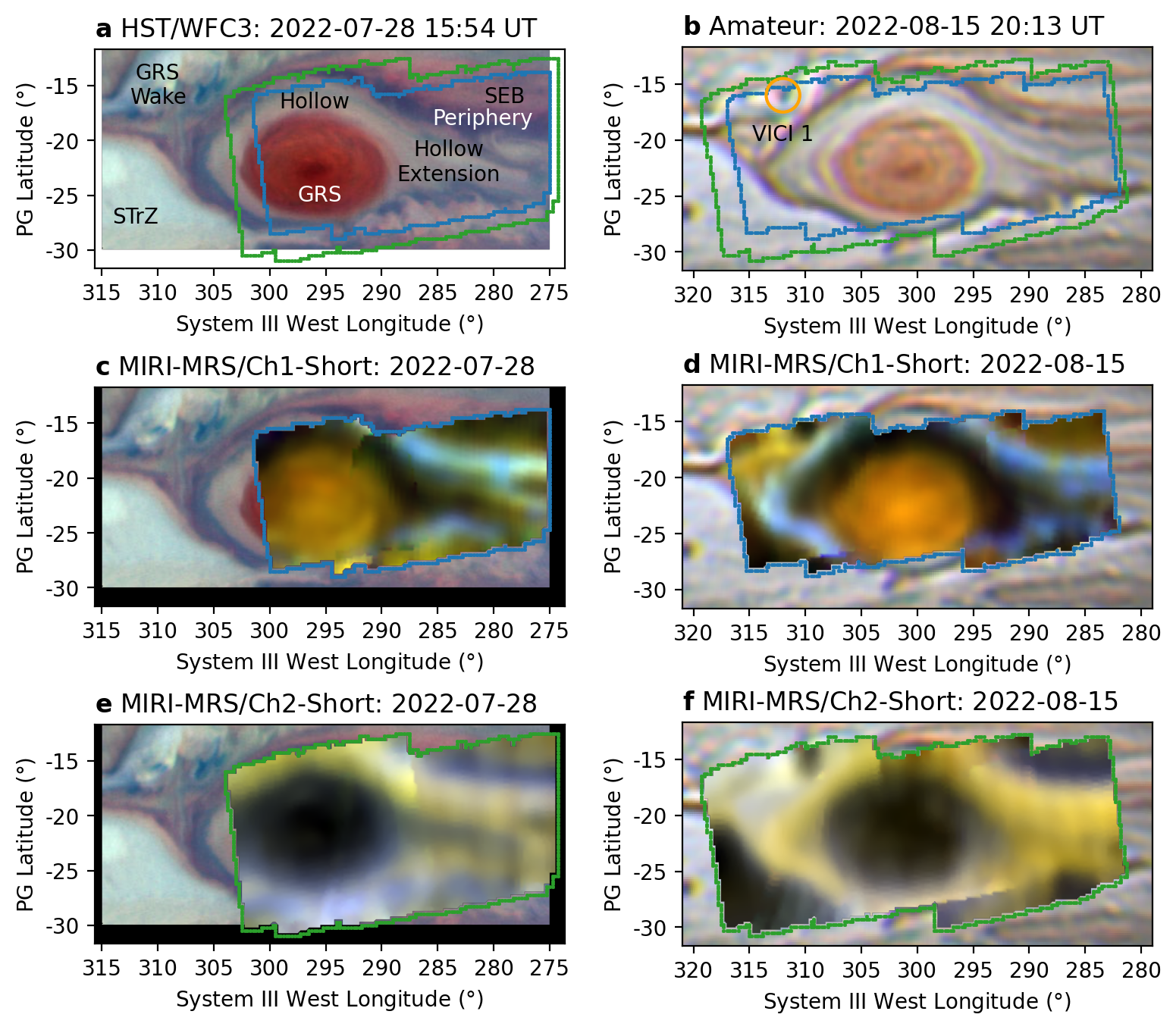}
        \caption{Comparison of False-colour JWST/MIRI data from channels 1 and 2 to visible context data from (a) Hubble (R=631 nm, G=502 nm, B=395 nm) and (b) Ground-Based amateur observations \protect\cite{ALPO_data}. The left-hand column displays data from 2022-07-28 while the right-hand column displays data from 2022-08-15. Differences in colour between the visible observations are due to different instruments and post-processing techniques being used. The size of the MIRI FOV is indicated by the blue and green lines for channels 1 and 2 respectively. Major atmospheric features are labelled on these plots. `Periphery' refers to the dark cloud-free lane of depleted opacity surrounding the GRS hollow. (c) and (d) False-colour images of the ch1-short GRS MIRI data with the relevant visual context image as the background. For both MIRI images; B = 5.40 µm, G = 5.50 µm and R = 5.60 µm. The red band contains reflected sunlight from within a deep NH$_3$ absorption band, indicating strong aerosol reflection rather than temperature. The green and blue bands are dominated by thermal emission, darker blue and green components outside the GRS therefore correspond to thick aerosol layers outside the GRS. (e) and (f) False-colour images of the ch2-short GRS MIRI data. For both; B = 8.62 µm, G = 8.57 µm and R = 8.56 µm. Beyond 7.30 µm, the jovian spectrum is dominated by thermal emission, darker regions correspond to regions of high aerosol opacity and cooler temperatures. All frames have been shifted to be centred on the GRS longitudinal position in July and August, taken to be (296.0$\degree$W, 22.5$\degree$S) and (301.0$\degree$W, 22.5$\degree$S) respectively. Subtle changes in colour between epochs are due to the different observing geometries of the observations.}
        \label{jup_rgb_troposphere}
    \end{figure}

\section{Data and calibration}
\label{data_methods}

    \subsection{Observations}

        \subsubsection{MIRI}
            \label{data_miri}

            Fig. \ref{jup_rgb_troposphere} displays mosaics of the GRS observations alongside contextual data from Hubble and amateur observations, taken on 2022-07-28 and 2022-08-15. JWST/MIRI observed this region using the Medium Resolution Spectrometer (MRS) \cite{g} as part of Guaranteed-Time Observations awarded to Heidi B. Hammel (cycle 1 - GTO 1246, PI: Leigh N. Fletcher). The MRS provides medium-resolution spectroscopy with a spectral range of 4.9 - 27.9 µm. Spectral resolution varies from R = 3,320 at 4.9 µm to R = 1,330 at 27.9 µm \cite{law_2023drizzle_paper}. MIRI achieves this using a set of 4 integral field units (channels 1 - 4), each of which cover a different portion of the MRS spectral range. All 4 channels are observed simultaneously, but only a third of the available spectral range is observed at any given time. To record the full spectral range, a dichroic grating wheel with 3 different grating settings or sub-bands (short/A, medium/B and long/C) is used to provide full spectral coverage \cite{20}. The time taken for this wheel to rotate to a new sub-band resulted in a short delay on the order of minutes between adjacent parts of the spectrum. The full dataset consisted of 12 MIRI `bands' spanning the full spectral range of the instrument.
            \par
            The FOV, spatial pixel (spaxel) size and slice width for each channel also varied. Channel 1 possesses the smallest FOV of 3.2 × 3.7 arcsec, but also the smallest spaxel size and slice width of 0.196 arcsec and 0.176 arcsec respectively. Channel 4 had the largest FOV of 6.9 × 7.9 arcsec and the largest spaxel size and slice width of 0.273 arcsec and 0.645 arcsec respectively. The diffraction-limited spatial resolution of the instrument at 9.0 µm at the time of the observations was 0.4 arcsec, corresponding to a distance of 1,090 km on Jupiter or a longitudinal distance of 0.9$\degree$ at the latitude of the GRS. Due to the small size of the instrument FOV compared to the jovian angular diameter of 45 arcsec, three observations were acquired for each epoch targeting the area west of the GRS, the centre and the east respectively. Table \ref{observation_table} displays a list of the observations and the longitudes targeted by each observation. The longitude of the GRS was estimated based on longitude drift rates determined by the OPAL programme \cite{simon_opal_paper} and amateur observer submissions to the Planetary Virtual Observatory and Laboratory (PVOL) \cite{hueso_2018pvol_paper}. MIRI uses a non-destructive up-the-ramp readout process \cite{p} where each exposure is split into a series of 8 smaller exposures known as `integrations'. The brightness is reset at the end of each integration to mitigate the risk of saturation. These integrations are further subdivided into four 2.8 s sub-exposures known as `groups'. The effective exposure time used for each band and dither was 89 s with buffer time around each exposure to allow the grating wheel to rotate to the next setting, this gave each observation a total time of 433 s. Note that 4 groups per integration was shorter than the recommended (and tested) number of groups by STScI of 5. This was chosen in order to both minimise saturation in the data as well as test the calibration pipeline's ability to work with data with an abnormally low number of groups. No problems were encountered by doing this. It was possible to further remove groups in post-processing to allow for even shorter integration times, enabling the recovery of data in saturated regions of the spectrum as shown in Section \ref{data_pipeline}. A 4-point dither pattern optimised for extended sources was used for each tile to improve spatial sampling in post-processing.
            \par
            The sequence of observations in July was chosen to account for Jupiter's rapid rotation in an attempt to capture each mosaic tile as close to the centre of the jovian disc as possible. The eastern tile was the first to be observed on 2022-07-28 (04:00 - 04:59UT), followed by the central tile (05:21 - 06:20UT). Guide stars were observed during the acquisition of each tile by the JWST Fine Guidance Sensor to continually update the observatory pointing and attitude control. A different guide star was observed for each tile. The first of these stars was acquired prior to the eastern tile observation (03:45 - 03:58UT). The second was acquired between the eastern and central tile observation (05:06 - 05:19UT). The third however, scheduled after the central tile observation was delayed due to a failure to acquire the guide star as a result of unresolved observatory pointing errors \cite{q}. This caused the GRS to rotate onto the far side of the planet before the western tile could be captured, resulting in the observation being missed. The background exposure (06:51 - 07:01UT) was therefore taken instead, targeting a region 90 arcsec from the disc of Jupiter to characterise any instrumental artefacts. Instructions were uploaded to the observatory to repeat the entire sequence of observations in August. This time the sequence began with the central tile (2022-08-14 23:04UT - 2022-08-15 00:03UT), followed by the western tile (2022-08-15 00:37 - 01:35UT) with the observatory then waiting a full jovian rotation for the eastern tile to return into the FOV (2022-08-15 09:33 - 09:38UT). A further guide star acquisition issue prior to the eastern tile observation resulted in the GRS rotating out of the FOV a second time before observations were complete. As a result only the first two dither positions and one grating setting (A) were observed for the eastern tile in August. 
            \par
            Despite the background exposures captured in July being displaced 90 arcsec from the disc of Jupiter, there was still considerable jovian stray-light present. However, the mean Signal-to-Noise ratio (S/N) in much of the 7.30 - 10.75 µm range exceeded 220, negating the need for this background exposure. A decision was therefore made to not use these background frames in the calibration of the data (to prevent contamination by jovian stray light) and a planned background exposure in August was not attempted. Finally, the jovian planetographic solution from NAIF \cite{acton_1996NAIF_paper, acton_2018naif_paper} was used to assign system III West longitude and latitude to the data, assuming solid body rotation at the system III rate and neglecting the tropospheric winds.

        \subsubsection{NIRCAM}
            \label{data_nircam}

            \begin{figure}
                \centering
                \includegraphics[width=\textwidth]{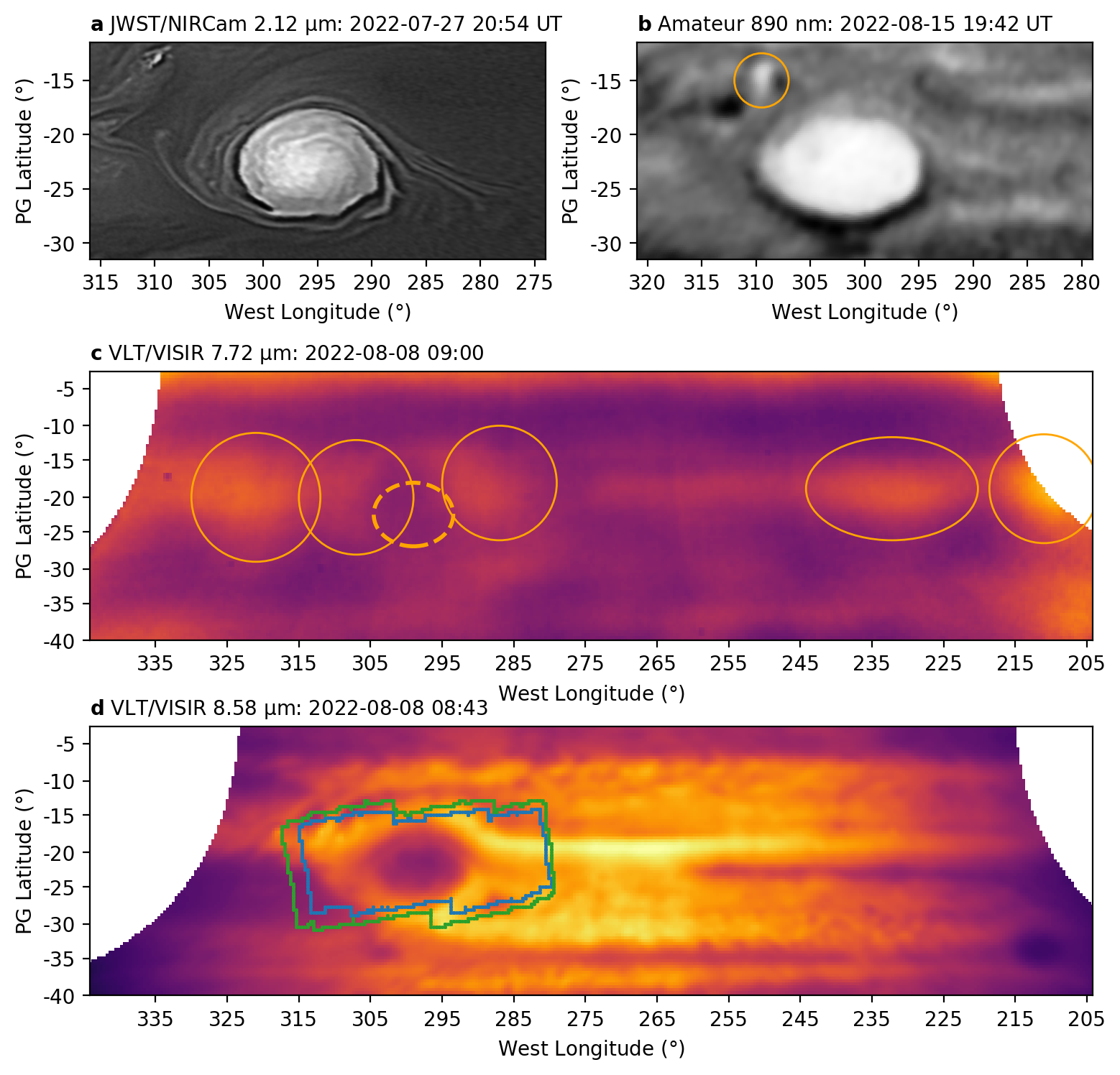}
                \caption{(a) JWST/NIRCam F212N image of the GRS and surroundings obtained on the same date as the JWST/MIRI July observations. (b) Ground-based 890 nm amateur observations taken on 2022-08-15 \protect\cite{ALPO_data}. This wavelength is within a region of strong CH$_4$ absorption. As a result, any bright features within this image are likely to be sunlight reflecting off aerosol formations that are elevated in altitude. The location of a storm north-west of the GRS is indicated by the orange circle. This is further discussed in Section \ref{NEMESIS}. The FOV for both images is the same as for Fig. \ref{jup_rgb_troposphere}. (c) VLT/VISIR observations taken using the J7.9 filter (7.72 µm). Contribution in this spectral range is from the jovian stratosphere. The position of the GRS below is indicated by the dashed orange circle. A series of anomalous hot-spots are highlighted by the solid orange circles, these are further investigated in Sections \ref{NEMESIS} and \ref{results_temp}. (d) VLT/VISIR observations taken using the PAH1 filter (8.58 µm). Contribution is from the jovian troposphere. As in Fig. \ref{jup_rgb_troposphere}, the size of the MIRI August epoch channel 1 and 2 FOVs are indicated by the blue and green lines respectively. These have been shifted by 2$\degree$ east to be centred on the GRS. The GRS can be seen centred within the MIRI FOV as a region of anomalously low temperature compared to the surroundings.}
                \label{general_context_figures}
            \end{figure}

            The entire jovian disc, including the GRS was observed using the JWST NIRCam instrument \cite{r} as part of the Early Release Science programme (ERS 1373, PIs: Imke de Pater; Thierry Fouchet). NIRCam is designed to provide near-infrared imaging of a target in a 9.7 arcmin$^{2}$ FOV, with an available spectral range of 0.6 - 5.0 µm. This is achieved by using a beam-splitting dichroic to generate 2 simultaneous images in a Short-Wavelength Channel (SWC: 0.6 - 2.3 µm) and a Long-Wavelength Channel (LWC: 2.4 - 5.0 µm). A series of 29 filters are available spanning the full spectral range. These are selected using a grating wheel similar to MIRI MRS. The aperture of the LWC consists of a 2.2 arcminute wide detector while the SWC utilises four 64 arcsec wide detectors arranged in a square to approximately match the FOV of the LWC. There is a separation of 5 arcseconds between each of the four SWC detectors, therefore a dither pattern was chosen for these observations that enabled recovery of the gaps in post-processing. An additional LWC and SWC aperture displaced 42 arcsec from the first is available possessing the same properties as the first. However, only one aperture was required to capture the whole disc of Jupiter. 
            \par
            The F212N filter was used for this analysis with a central wavelength of 2.120 µm and a bandwidth of 0.027 µm. An effective exposure time of 53.7 s was used for this study. The resulting diffraction-limited spatial resolution of the instrument/filter combination in this case was 0.08 arcsec, corresponding to a distance of 260 km on Jupiter or a longitudinal difference of 0.2$\degree$. Two sets of observations were made on 2022-07-28 at 10:55UT and 20:58UT respectively. The timings were determined to allow approximately a whole jovian day (9.93 hrs) to pass between observations in order to track wind velocities. F212N was chosen as its position within a spectral region of strong H$_{2}$ Collision Induced Absorption (CIA) mitigated the risk of saturation. The Particle Image Correlation Velocimetry 3 (PICV3) software \cite{sss} was used to derive the wind velocities within the GRS over the 2 sets of observations at a presumed altitude of 240 mbar \cite{hueso_2023nircam_paper}. These velocities are compared to the thermal wind velocities derived using the JWST/MIRI data in Section \ref{thermal_winds}. An analysis of winds in Jupiter's Equatorial Zone (EZ) using these data and a description of the calibration and analysis of the NIRCam data were published in \citeA{hueso_2023nircam_paper}. The data used in this study can be seen in Fig. \ref{general_context_figures}a.
            \par
            Visual and near-infrared context imagery at 890 nm on 2022-08-15 were provided by amateur observers \cite{ALPO_data}. Longitude and latitude grids were assigned to both the amateur and the NIRCam observations using the PlanetMapper software \cite{king_2023planetmapper_paper} and can be seen in Fig. \ref{jup_rgb_troposphere}b, \ref{general_context_figures}a and b.

        \subsubsection{VLT/VISIR}
            \label{data_visir}

            Between the two dates of JWST/MIRI observations, Jupiter was further observed by the Very Large Telescope Imager and Spectrometer (VLT/VISIR) as part of programme 108.223F.001, providing ground-based infrared observations in support of Juno \cite{fletcher_2017vltneb_paper, fletcher_2017vltseb_paper, fletcher_2018vltjiram_paper, antunano_2020visirez_paper, bardet_2024visirjupiter_paper}. This is optimised to provide diffraction-limited mid-infrared imaging in the M-band (5 µm), the N-band (8 - 13 µm) and the Q-band (17-20 µm) \cite{lagage_2004visir_paper}. The imaging is generally performed in a 38 $\times$ 38 arcsec FOV with a pixel scale of 0.045 arcsec. Considerable background noise is induced in the data by the black-body emission of the Earth's atmosphere. This is mitigated by rapidly moving the telescope secondary mirror on and off the target in a motion known as `chopping', alternately capturing the data and the background. The latter of these can then be subtracted from the former to remove the rapidly fluctuating background emission. However, due to the optical path of the sets of chopped observations being different, residual background noise can linger in the FOV. This is removed by `nodding'. The same chopped observations are performed off the target to generate a flat-field of this remaining noise, which can then be subtracted from the data.
            \par
            The J7.9 (central wavelength: 7.72 µm, filter width: 0.56 µm) and PAH1 (central wavelength: 8.58 µm, filter width: 0.41 µm) filters were used for this study. The resulting diffraction-limited spatial resolution of the instrument at 8.15 µm (between the two filters) was 0.25 arcsec, corresponding to a distance of 728 km or a longitudinal difference of 0.6$\degree$. Following the chopping and nodding routine above, the images were also assigned longitude and latitude grids using PlanetMapper and were subsequently regridded onto identical rectangular longitude/latitude grids with a pixel scale of 0.4$\degree$ to ensure the J7.9 data would not be undersampled. This data can be seen in Fig. \ref{general_context_figures}c and d.

        \subsubsection{Hubble}
            \label{data_hubble}
        
            The entire jovian disc was also observed at 631 nm using the Hubble Space Telescope's Wide Field Camera 3 (HST/WFC3) \cite{t}. This provides UV, Visible and near-infrared imaging in the 200 - 1,700 nm spectral range. It achieves this using two channels; UVIS (200 - 1,000 nm - 160 $\times$ 160 arcsec FOV) and NIR (850 - 1,700 nm - 123 $\times$ 137 arcsec FOV) alongside a complement of 77 filters to provide the full spectral range. The UVIS/F631N filter, centred on 630.6 nm with a bandwidth of 5.4 nm was chosen as it is sensitive to the wind field at the top of the tropospheric cloud layers. However, the altitude range was loosely constrained to the 600 - 500 mbar range \cite{west_2004jupitercloud_book} and is a topic that will be further discussed in Section \ref{thermal_winds}. The diffraction-limited spatial resolution at this wavelength was 0.07 arcsec, corresponding to a distance of 210 km on Jupiter or a longitudinal difference of 0.2$\degree$. This was comparable to the NIRCam F212N observations. An observation made on 2022-07-28 was used for spatial context for the July MIRI observations and can be seen in Fig. \ref{jup_rgb_troposphere}.
            \par
            GRS wind velocities were inferred using a median of 3 Hubble wind measurements made on 2020-09-20, 2021-09-04 and 2023-01-06. For each of the dates, four 4 second exposures spanning approximately 11 hrs were used to determine the GRS wind velocities using the Advection Corrected Correlation Image Velocimetry technique (ACCIV) \cite{asay-davis_2009grswind_paper}. Observations taken closer to the dates of the MIRI data on 2022-11-12 were not used due to the two pairs of exposures only being separated by 95 minutes, increasing the resulting uncertainty on the winds by a factor of 3-4.
            \par
            The derived median velocity field from these observations were interpolated onto the same lon/lat grid as the MIRI data before being used as the prior wind field for the derivation of thermal winds from both the July and August MIRI data. The results of this can be seen in Section \ref{thermal_winds}. The absence of Hubble observations on the date of the August epoch resulted in our use of visual observations taken by amateur observers \cite{ALPO_data}.
        
        \begin{table}
            \centering
            \begin{tabular}{c c c c c c c}
                \hline
                Instrument & Mode & Date Start (UTC) & Duration & Lon & Lat & $\theta_E$ \\
                \hline
                2020 HST/WFC3 I & F631N & 2020-09-20 02:59:28 & 00:00:04 & 78.4 & 0.0 & 0.0\\
                2020 HST/WFC3 II & F631N & 2020-09-20 03:12:41 & 00:00:04 & 78.4 & 0.0 & 0.0\\
                2020 HST/WFC3 III & F631N & 2020-09-20 12:30:58 & 00:00:04 & 78.4 & 0.0 & 0.0\\
                2020 HST/WFC3 IV & F631N & 2020-09-20 12:57:42 & 00:00:04 & 78.4 & 0.0 & 0.0\\
                2021 HST/WFC3 I & F631N & 2021-09-04 07:44:15 & 00:00:04 & 189.1 & 0.0 & 0.0\\
                2021 HST/WFC3 II & F631N & 2021-09-04 08:42:54 & 00:00:04 & 189.1 & 0.0 & 0.0\\
                2021 HST/WFC3 III & F631N & 2021-09-04 17:16:35 & 00:00:04 & 189.1 & 0.0 & 0.0\\
                2021 HST/WFC3 IV & F631N & 2021-09-04 18:19:00 & 00:00:04 & 189.1 & 0.0 & 0.0\\
                JWST/NIRCam I & F212N & 2022-07-27 10:51:40 & 00:08:35 & 342.9 & 0.0 & 0.0\\
                JWST/NIRCam II & F212N & 2022-07-27 20:54:20 & 00:08:25 & 347.3 & 0.0 & 0.0\\
                JWST/MIRI East I & MRS & 2022-07-28 04:00:41 & 00:58:55 & 286.0 & 22.5 & 35.5\\
                JWST/MIRI Cent I & MRS & 2022-07-28 05:21:43 & 00:58:36 & 296.0 & 22.5 & 29.6\\
                JWST/MIRI West I & MRS & N/A & N/A & 306.0 & 22.5 & N/A\\
                JWST/MIRI BKG I & MRS & 2022-07-28 06:51:35 & 00:09:48 & N/A & N/A & N/A\\
                2022 HST/WFC3 I & F631N & 2022-07-28 15:45:06 & 00:00:04 & 296.0 & 0.0 & 0.0\\
                2022 HST/WFC3 II & F502N & 2022-07-28 15:59:57 & 00:00:04 & 296.0 & 0.0 & 0.0\\
                2022 HST/WFC3 III & F395N & 2022-07-28 16:03:19 & 00:00:09 & 296.0 & 0.0 & 0.0\\
                VLT/VISIR & PAH1 & 2022-08-08 08:42:45 & 00:00:45 & 0.0 & 0.0 & 0.0\\
                VLT/VISIR & J7.9 & 2022-08-08 09:00:04 & 00:01:38 & 0.0 & 0.0 & 0.0\\
                JWST/MIRI Cent II & MRS & 2022-08-14 23:04:31 & 00:58:36 & 301.0 & 22.5 & 37.6\\
                JWST/MIRI West II & MRS & 2022-08-15 00:37:21 & 00:58:39 & 311.0 & 22.5 & 31.4\\
                JWST/MIRI East II & MRS & 2022-08-15 09:33:38 & 00:06:40 & 291.0 & 22.5 & 27.7\\
                2023 HST/WFC3 I & F631N & 2023-01-06 07:53:44 & 00:00:04 & 347.6 & 0.0 & 0.0\\
                2023 HST/WFC3 II & F631N & 2023-01-06 08:31:07 & 00:00:04 & 347.6 & 0.0 & 0.0\\
                2023 HST/WFC3 III & F631N & 2023-01-06 18:54:51 & 00:00:04 & 347.6 & 0.0 & 0.0\\
                2023 HST/WFC3 IV & F631N & 2023-01-06 19:31:35 & 00:00:04 & 347.6 & 0.0 & 0.0\\
                \hline
            \end{tabular}
            \caption{Summary of all observations listing (where applicable) either the filter used or the observing mode and the start and end times of the observations. The JWST data times exclude the guide star acquisition observation performed between each tile. Where necessary, the start and end times have been rounded to the nearest second. Longitude (Lon) and Latitude (Lat) are planetographic and the longitude is System III west. Both are in degrees. Emission angle ($\theta_E$) is also in degrees. The JWST/MIRI Background (BKG) observation was positioned 90 arcsec north of the centre of the jovian disc.}
            \label{observation_table}
        \end{table}

    \subsection{JWST pipeline calibration}
        \label{data_pipeline}
    
        The raw stage-0 \verb|UNCAL| data were calibrated using the Python JWST calibration pipeline (version 
        1.11.4) using a Calibration Reference Data System (CRDS) context defined by \verb|jwst_1112.pmap| \cite{34}. This was applied to the stage-0 \verb|UNCAL| data cubes available from the Mikulski Archive for Space Telescopes (MAST). The pipeline consists of 3 stages; Stage 1 corrects the individual integration ramps for saturation, dark current and jumps caused by cosmic rays hits (by removing the affected groups) before assembling these ramps into a single integration slope for each spaxel. Stage 2 applies flux calibration, wavelength calibration and corrects for straylight, flat-field effects and fringing. Stage 3 applies a final outlier-detection routine to remove cosmic ray hits not removed in Stage 1 before drizzling the data onto a rectangular grid to generate 3D cubes of the data where two axes are spatial and one spectral. Stage 1 is the same regardless of the instrument used. However, starting from Stages 2 and 3, the pipeline becomes more specific to the instrument used and the type of data obtained. The \verb|Spec2Pipeline| and \verb|Spec3Pipeline| versions of Stage 2 and Stage 3 were used to calibrate the MIRI/MRS data respectively. The Doppler shift caused by the 24 km s$^{-1}$ relative motion between Jupiter and JWST was accounted for in Stage 2 of the calibration pipeline. Therefore the only residual Doppler-shift in the data was due to planetary rotation. This was estimated to typically be ±1 km s$^{-1}$ or less, corresponding to a shift of 0.02 - 0.09 nm across the MIRI/MRS spectral range. This shift was negligible compared to the wave steps of 0.80 nm and 1.30 nm in channels 1 and 2 respectively. Also, coherent reflections inside the instrument generated a series of brightness-dependent periodic fringes \cite{55}. These were mitigated by the \verb|FringeStep| and the \verb|ResidualFringeStep| steps, both run during stage 2 of the calibration pipeline. However, due to the abundance of molecular lines in the spectrum compared to other targets, the fringe correction for wavelengths longer than 10.75 µm was limited. We therefore decided that the data beyond this threshold should not be used until a better fringe correction became available.
        \par        
        Considerable post-processing was required for each individual dither position to render the MRS data usable. Therefore only the stage 0 data were downloaded from MAST and the three pipeline stages were run locally. Our custom post-processing pipeline navigated the data files, significantly reduced the impact of saturation, corrected residual flat-field effects and removed anomalous spikes from the data. It was originally developed for similar MIRI/MRS observations of Saturn \cite{fletcher_jwstsaturn_paper} and is fully described in \citeA{king_jwstcalibration_paper}.
        \par
\begin{figure}
            \centering
            \includegraphics[width=1.0\textwidth]{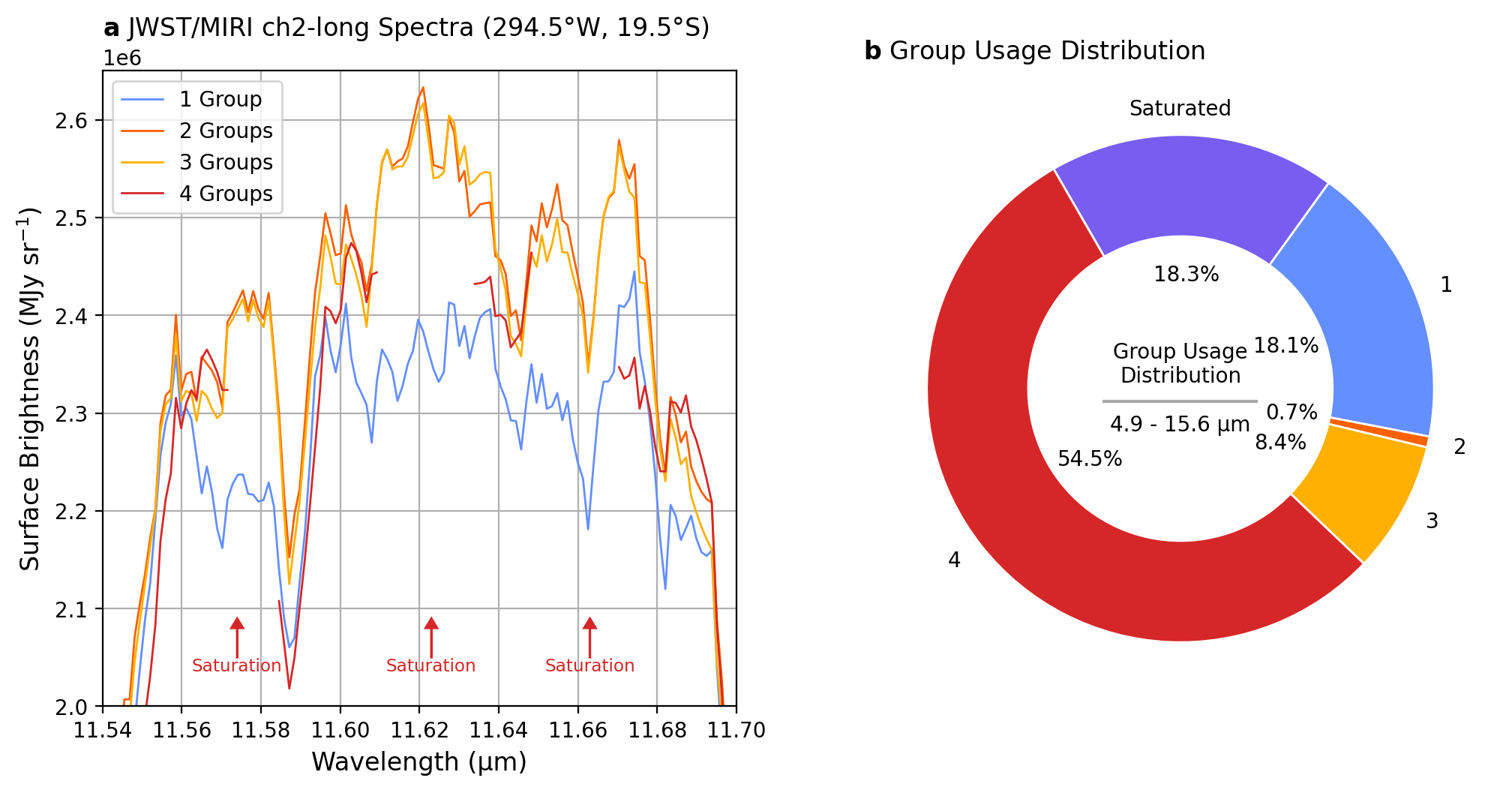}
            \caption{(a) Example spectra from (294.5$\degree$W, 19.5$\degree$S) within ch2-long displaying the effect of removing groups from the stage-0 UNCAL data before running the JWST calibration pipeline. The spectrum with 4 groups experiences 3 areas of saturation centred around 11.58 µm, 11.62 µm and 11.66 µm respectively. These regions are recovered in the spectrum using 3 groups. However, reduced S/N results in severe deviation in the 1-group spectrum. Note the similarity of the 2 and 3 group data. (b) Distribution of the number of groups used by each spaxel in the ch1-short to ch3-medium range. The majority continued to use 4 groups with saturation only becoming a major issue past 11.0 µm (ch2-long). Despite having access to 1-group data, 18.3\% of spaxels were still unrecoverable. The largest number of groups was used in each case in order to desaturate the data while maintaining the highest S/N possible.}
            \label{sat_figure}
        \end{figure}

         Due to the brightness of Jupiter, the data are saturated beyond 11 µm as well as the brighter regions of ch1-short. Removal of groups was shown to recover data in these regions up to 15.57 µm (end of ch3-medium). However, removal of too many groups degraded the Signal-to-Noise (S/N) to the point where the data were unusable. The desaturation routine presented in \citeA{fletcher_jwstsaturn_paper, king_jwstcalibration_paper} was used to recover the affected parts of the spectrum for each spaxel by replacing the saturated spectra with unsaturated data containing as many groups as possible without resaturating. Fig \ref{sat_figure} displays an example of the effect of removing groups from the spectrum as well as the distribution of groups used by each spaxel up to 15.57 µm (end of ch3-medium). The majority of the spectrum still utilised 4 groups. However, 27.2\% of the data used 3 groups or fewer and 18.3\% remained saturated despite the availability of 1-group data. For comparison, $<$ 10\% of the spaxels in MIRI data for Saturn (cycle 1 - GTO 1247) \cite{fletcher_jwstsaturn_paper} contained less than the maximum number of groups. Beyond ch3-medium, the data remained saturated even with 1 group and is therefore currently unusable.

        The \verb|CubeBuildStep| in stage 3 of the JWST calibration pipeline estimates spectral uncertainty, based both on the variance observed within the `drizzle' technique used to generate the final data cubes and observation noise (which decreases for higher S/N regions of the data). This is then output as the `ERR' backplane of the data. There is currently a known issue where these uncertainties are underestimated, sometimes by more than a factor of 10 \cite{law_2023drizzle_paper}. For the purpose of analysing this data, the uncertainties were multiplied by a factor ranging from 4-8 across the entire wavelength range, retaining the shape of the uncertainties but also obtaining a reduced goodness-of-fit ($\chi^2$) of approximately one in the atmospheric retrievals, equivalent to adding retrieval uncertainty into the process \cite{irwin_nemesis_paper}. 
        \par

        A number of non-physical stripes were also observed in the original 2022 stage 3 data that were determined to be the product of wavelength-calibration offsets within the IFU FOV. The magnitude of these offsets were derived by comparing the wavelength positions of molecular features in data from both Jupiter (GTO 1246) and Saturn (GTO 1247) to models generated using NEMESIS \cite{irwin_nemesis_paper}. These offsets were fitted across the wavelength range of each band using a Savitzky-Golay filter to produce the FLighT-5 (FLT-5) wavelength solution, fully described in \citeA{argyriou_wavecal_paper}. This solution is applied during the \verb|AssignWcsStep| as part of stage 2 of the standard JWST calibration pipeline, addressing this issue for all new MIRI/MRS data.
        \par

        After application of the FLT-5 wavelength solution and desaturation, there were still a significant number of wavelength-dependent flat-field effects. Again, this is a known pipeline issue and is currently under investigation but is likely to be associated with the construction of the cube from the dispersed spectra. \citeA{fletcher_jwstsaturn_paper, king_jwstcalibration_paper} derived flat-field solutions for each individual wavelength using JWST/MIRI observations of Saturn by comparing differences between similar dither positions within an oval aperture, the average of the unique flat-field solutions from each tile were successfully used to calibrate the Saturn data. However, these flat-fields could not remove the artefacts seen in the Jupiter data. Therefore, we derived similar mean flat-fields using the Jupiter data. A circular aperture with a radius equal to half the average latitude distance between spaxels was used this time due to Jupiter being equally variable in the meridional and zonal directions. Although the data quality improved, it did not fully address the flat-fielding problem. The results of applying these first two generations of flat-fields can be seen in Supplementary Figure S1. New `self' flat-fields were therefore derived using just the 4 dither positions from each tile. This produced a unique flat-field solution for each tile. Applying these to the data as seen in Supplementary Figure S2 resulted in considerable improvement to the data quality, with previously obscured atmospheric structure becoming visible. Mean standard deviation ($\sigma$) across the ch2-short wavelength range at (295.5$\degree$W, 20.5$\degree$S) for the Saturn Corrected, Jupiter corrected and Self-Corrected data as shown in Supplementary Figure S3 was; 1.17$\sigma$, 1.84$\sigma$ and 1.65$\sigma$ respectively. However, accounting for the underestimated pipeline uncertainties, the difference in surface brightness between the flat-field corrected and the pipeline output data is typically a factor of 0.2 - 0.5 smaller than the spectral uncertainties and was therefore ignored. One possible reason for Jupiter and Saturn requiring different flat-fields is that the flat-fielding effects experienced by MIRI/MRS may be brightness dependent. Each tile of the jovian observations also required a unique flat-field, possibly due to the changing emission angle causing subtle changes in illumination. Finally, a despiking routine outlined in \citeA{king_jwstcalibration_paper} was also applied to the data to remove anomalous spikes generated by cosmic rays and noise not removed by the standard calibration pipeline.
        \par
        Once the post-pipeline calibration steps had been carried out, the four dither positions (two for August-East) for each tile were combined and re-gridded onto a regular west-longitude and planetographic latitude grid with a scale of 0.5$\degree$ $\times$ 0.5$\degree$ using a Nyquist sampling technique. This determined the mean-combination of all data within a square bin 1.5$\degree$ wide surrounding each spaxel in the new grid. The 0.5$\degree$ scale was chosen to avoid undersampling the data based on the diffraction-limited spatial resolution calculated in Section \ref{data_miri}. The results of this were the input data for the atmospheric retrievals, presented in Sections \ref{NEMESIS} and \ref{results}.

\section{Radiative transfer modelling and thermal winds}
\label{NEMESIS}

    \begin{figure}
        \centering
        \includegraphics[width=\textwidth]{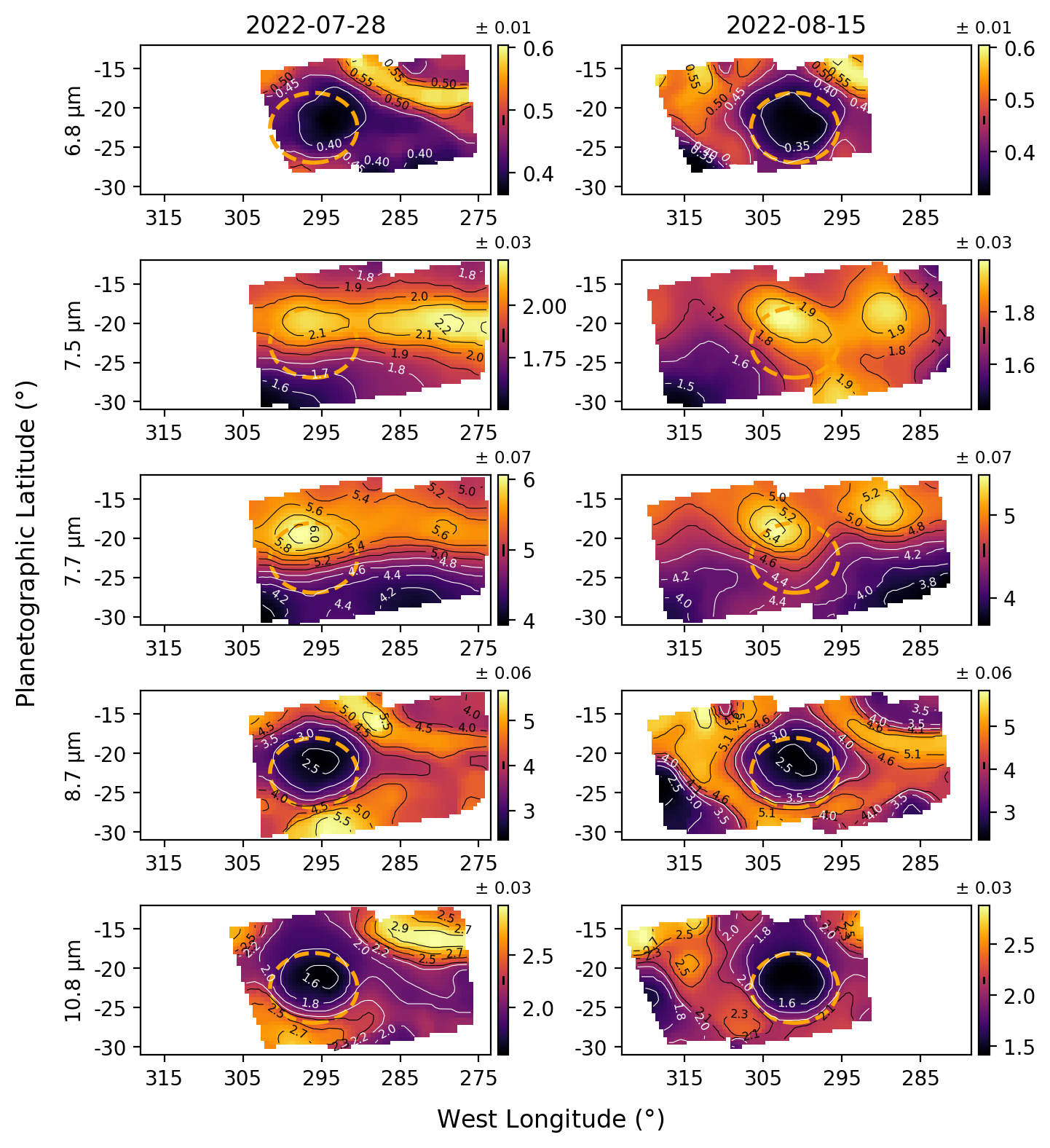}
        \caption{A series of spatial maps of radiance (µW cm$^{2}$ sr$^{-1}$ µm$^{-1}$) for wavelengths in ch1-long to ch2-long. 6.8 µm, 8.7 µm and 10.8 µm probe the 800 mbar, 100 mbar and 400 mbar pressure levels respectively. Both the GRS and the storm located in the GRS wake at (310$\degree$W, 15$\degree$S) appear dark at these tropospheric wavelengths compared to the surroundings due to the presence of cold temperature anomalies. 7.5 and 7.7 µm probe the 10 mbar and 1 mbar pressure levels respectively, corresponding to the jovian stratosphere. These two altitudes display a remarkable amount of spatial and temporal variation. Differences in brightness between the two epochs are likely due to different observing geometries. The median 1$\sigma$ uncertainties are displayed as error bars in each of the colour bars, with the numerical values displayed above the colour bars. 2D maps of these uncertainties can be seen in Supplementary Figure S4. The positions of the peak Hubble wind velocities are indicated by the dashed orange circle. The first column displays the July epoch uncertainties, the GRS is centred on (296.0$\degree$W, 22.5$\degree$S). The second column displays the August data, the GRS here is centred on (301.0$\degree$W, 22.5$\degree$S).}
        \label{TB_comp}
    \end{figure}

    Fig. \ref{TB_comp} displays brightness temperature maps (T$_b$) of the MIRI data for several wavelengths, probing a range of tropospheric and stratospheric altitudes. Fig. \ref{jup_rgb_troposphere} displays the Hubble and amateur ground-based observations of the GRS that took place on the same dates as the MIRI observations. The cold temperature anomaly of the GRS dominates wavelengths containing contribution from the troposphere. Meanwhile in the stratosphere, a series of hot-spots were observed at 7.7 µm that are further investigated in Section \ref{results_temp}. We were also fortunate to capture a small storm within this region in August seen at (310$\degree$W, 15$\degree$S) in both the 2022-08-15 MIRI and visible amateur observations. The near-infrared observations shown in Fig. \ref{general_context_figures} indicate a region of high albedo coincident with this storm at 890 nm, suggesting that the aerosol layers are elevated compared to the surroundings. This storm will be hereafter referred to as VIgorous Convective Instability 1 (VICI 1). It is possible that the upwelling air in this region is sufficient to allow detection of NH$_3$ ice with JWST/MIRI. This is further investigated in Section \ref{nh3_non_de}.
    \par
    We used the NEMESIS suite of radiative transfer and spectral inversion software \cite{irwin_nemesis_paper} to retrieve temperature and compositional distributions from the regridded MRS spectral cubes (Section \ref{data_pipeline}). The spectral database used for the radiative transfer modelling was primarily based on that used for Cassini retrievals \cite{j}, with NH$_3$ and PH$_3$ abundances with respect to altitude being inferred from the Jupiter flyby of Cassini/CIRS \cite{fletcher_2009ph3cirs_paper}. A series of updates were made to the abundances of AsH$_3$, GeH$_4$ and CO from \citeA{w}. A mean of C$_2$H$_2$ and C$_2$H$_6$ in the 30$\degree$N - 30$\degree$S region was used from \citeA{x}, also derived from Cassini CIRS data. In addition C$_2$H$_4$, C$_4$H$_2$, C$_3$H$_8$, CO$_2$, C$_6$H$_6$ and CH$_3$ were obtained from Jupiter Model C \cite{tt}. HCN was obtained from \citeA{ss}. The H$_2$ and He abundances were determined using data from the Galileo Probe \cite{aa} and the abundances of CH$_4$ and the two isotopologues; CH$_3$D and 13CH$_4$ were obtained from Voyager and Cassini CIRS observations with the CH$_4$ deep abundance of 1,810 ppm being based on the results of the Galileo probe \cite{n, bb}. Voigt broadening was assumed for all wavelengths with a line wing cut-off of 35 cm$^{-1}$. CIA of H$_2$ and He were derived based on the database discussed by \citeA{z}. The correlated-k method was utilised for this study to improve computational efficiency while still retaining accuracy. K-distributions were derived for each gas in the entire ch1-short to ch3-long spectral range. These distributions were determined for the level 3 pipeline wavelength grids with a spectral resolution derived from ground-testing \cite{20}. Note that updated in-flight measurements of the spectral resolving power \cite{y} have not yet been incorporated into the models. However, these have limited variations in channel 2, where the majority of this work is performed.
    \par
\begin{figure}
        \centering
        \includegraphics[width=\textwidth]{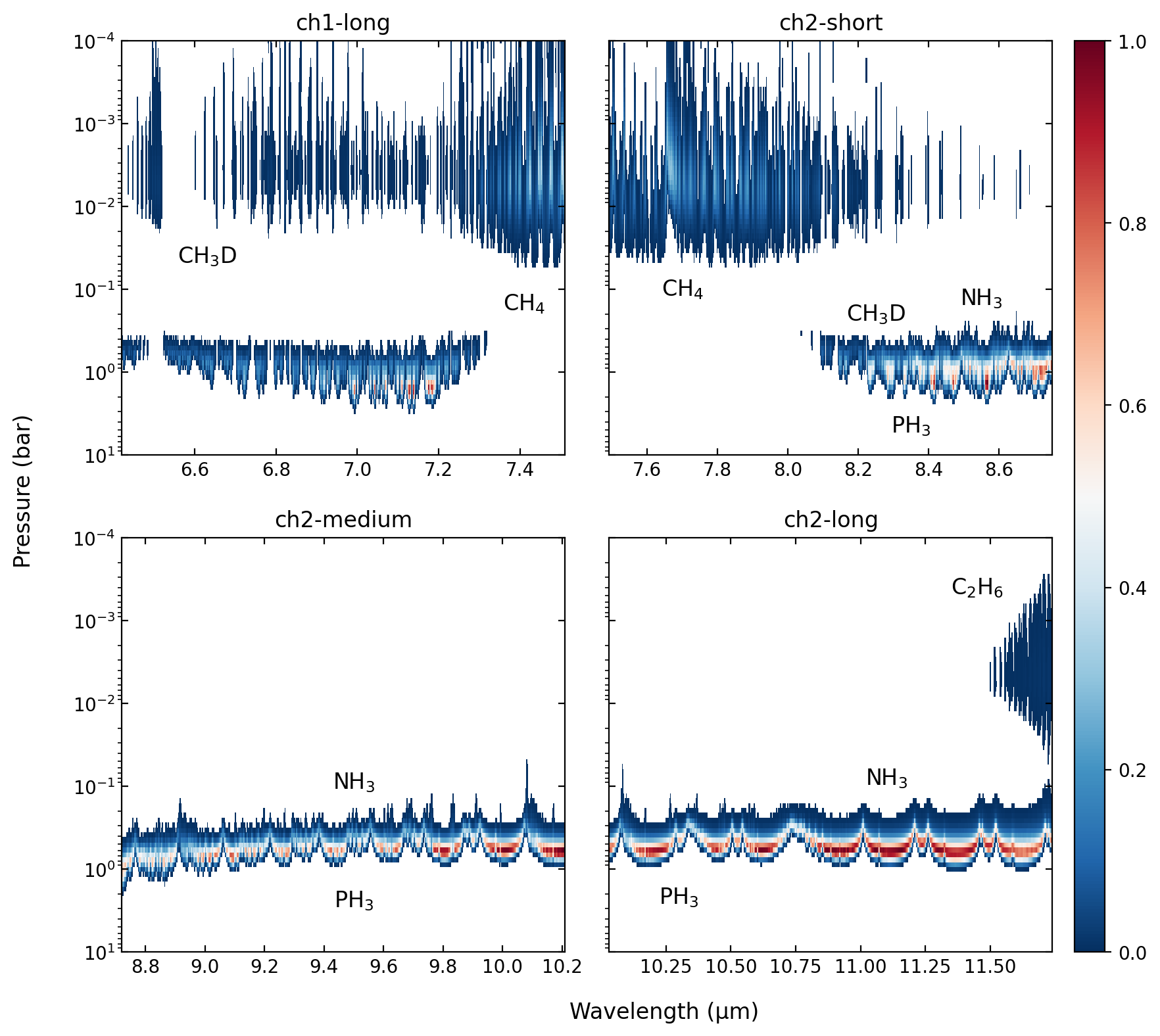}
        \caption{Cloud-free contribution functions (Normalised NEMESIS-derived Jacobians displaying the rate of change of spectral radiance with temperature for different layers of the jovian atmosphere), calculated assuming a nadir viewpoint on the equator for ch1-long to ch2-long. Stratospheric CH$_4$ emission increases the observed altitude within the emission feature, the latter part of the spectral range is dominated by tropospheric NH$_3$ and PH$_3$ absorption. The locations of major gaseous species are indicated for each band.}
        \label{contribution_plot}
    \end{figure}

    Due to the need to account for the scattering of reflected sunlight, the modelling and analysis of the 4.90 - 7.30 µm region will be the topic of a future paper. Beyond 7.30 µm, the effect of scattering due to reflected sunlight becomes negligible compared to jovian thermal emission. 7.30 - 8.10 µm is dominated by stratospheric CH$_4$ emission at altitudes above the 50 mbar pressure level. At longer wavelengths than this, the spectral shape becomes dominated by H$_2$ and He Collision-Induced Absorption (CIA) with other absorption and emission features overlaid on top of this curve. Until the end of ch2-long (11.70 µm) the majority of these features are associated with tropospheric NH$_3$ and PH$_3$ absorption. Fig. \ref{contribution_plot} displays the cloud-free contribution functions (indicating the approximate altitude sensitivity of the spectra) derived using NEMESIS. The locations of major molecular features are indicated on this plot. 

        3D temperatures, aerosol distribution, PH$_3$ and NH$_3$ maps were retrieved in the 7.30 - 10.75 µm spectral range (ch1-long - ch2-long), chosen to constrain both temperature and aerosol contribution from both the jovian troposphere and stratosphere as indicated by contribution plots shown in Fig. \ref{contribution_plot}.
        \par
        The prior temperature profile was a mean of the 1 - 1000 mbar profile in the 30$\degree$N - 30$\degree$S latitude range obtained by Cassini/CIRS \cite{fletcher_2009ph3cirs_paper}. The profile deeper than this was derived from a Dry Adiabatic Lapse Rate (DALR) modified to match the results for deep ($>$ 10 bar) temperature from the Galileo Probe Atmospheric Structure Instrument \cite{n}. Stratospheric temperatures were also obtained from Cassini/CIRS by \citeA{x}. Temperature was the only variable to be retrieved as a full continuous profile with altitude.
        \par
\begin{figure}
            \centering
            \includegraphics[width=\textwidth]{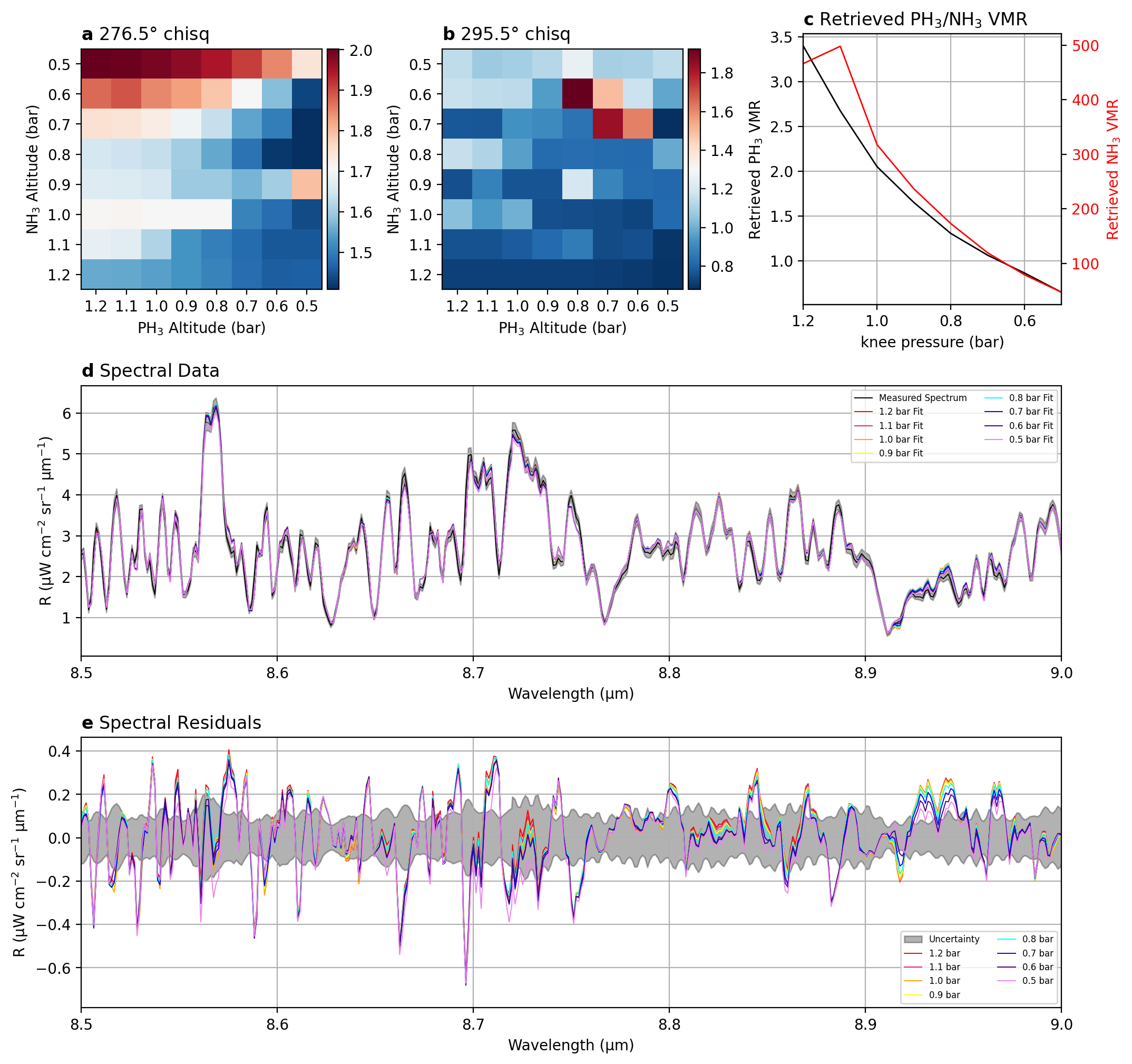}
            \caption{Results of the NEMESIS test retrievals to determine the NH$_3$ and PH$_3$ knee pressures to use in the retrievals due to the degeneracies between knee pressure and retrieved VMR. Data from July is displayed in each case and all data is centred on a latitude of 22.0$\degree$S. (a) Retrieved $\chi^2$ within the STrZ east of the GRS. A range of prior NH$_3$ and PH$_3$ knee pressures are shown. A gradient of $\chi^2$ can be seen across the plot indicating higher NH$_3$ and lower PH$_3$ knee pressures may be preferable. (b) A similar plot but for the centre of the GRS at the same latitude. Although the result is more ambiguous for this region, the trend observed in (a) is still visible here. (c) Values of PH$_3$ retrieved VMR (black) and NH$_3$ VMR (red), displaying the trend of decreasing retrieved VMR with increasing altitude of the knee pressure. A NH$_3$ knee pressure of 800 mbar retrieved VMR values consistent with Juno/MWR measurements of 200 ppm between 1-2 bar \protect\cite{li_2017mwrnh3_paper}. (d) Measured spectrum for (276.5$\degree$W, 22.0$\degree$S) alongside a range of spectra modelled using a constant NH$_3$ knee pressure of 0.8 bar and different assumptions for PH$_3$ knee pressure. (e) Corresponding spectral residuals for this position indicating that the best fit was achieved with a PH$_3$ knee pressure of 0.5 bar.}
            \label{nh3_ph3_knee_test} 
        \end{figure}

        NH$_3$ and PH$_3$ gaseous Volume Mixing Ratio (VMR) followed a prior value starting from a `knee' pressure and extending down to 10 bar (the lower boundary of the NEMESIS model atmosphere). At altitudes above the knee pressure, the VMR decreased with a Fractional Scale Height (FSH) consistent with Cassini/CIRS observations \cite{irwin_2004ph3cirs_paper}. VMR was set to drop exponentially to zero above the tropopause (100 mbar) \cite{baines_2019grschromophore_paper}. The prior NH$_3$ VMR was 200 ppm, consistent with 1-2 bar average Juno MicroWave Radiometer (MWR) measurements for the jovian southern hemisphere \cite{li_2017mwrnh3_paper}. For PH$_3$, the prior VMR was 0.5 ppm, consistent with Cassini/CIRS measurements \cite{fletcher_2009ph3cirs_paper}. A degeneracy between the knee pressures and the retrieved VMR required a series of test-retrievals, presented in Fig. \ref{nh3_ph3_knee_test} to determine the optimum values for knee pressure for both species. Pressures in the 1200 - 500 mbar range were tested, consistent with previous studies \cite{irwin_2004ph3cirs_paper, fletcher_2010grs_paper, fletcher_2016texes_paper, bjoraker_2018grswater_paper}. Knee pressures of 800 mbar and 500 mbar were chosen for NH$_3$ and PH$_3$ respectively as they provided retrieved VMRs that were consistent with previous observations while simultaneously maintaining good spectral fits. The prior values for these two species are summarised in Table \ref{prior_stats}.

\begin{figure}
            \centering
            \includegraphics[width=\textwidth]{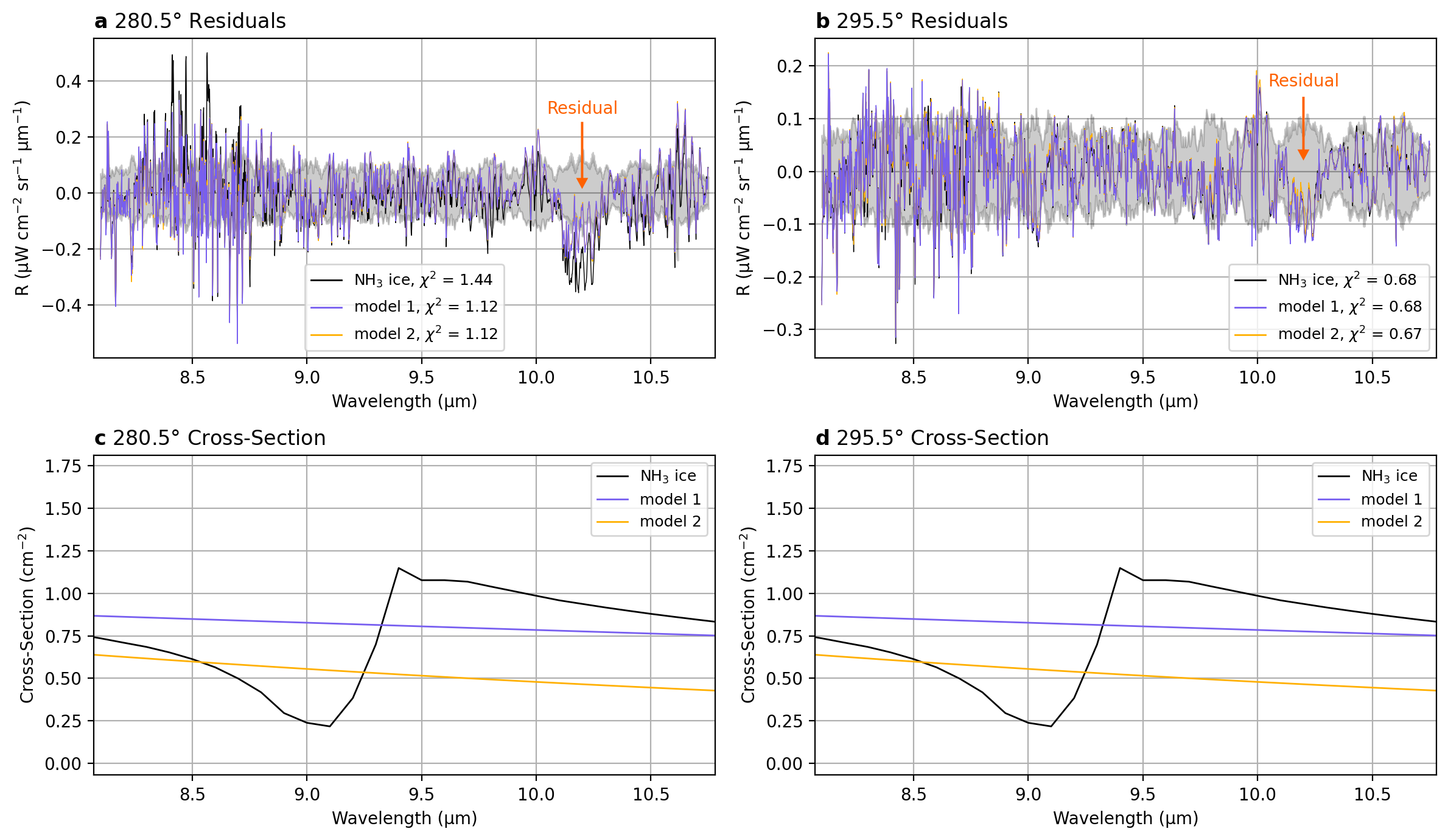}
            \caption{Results for the series of NEMESIS tests used to constrain optical properties of the aerosol layer. Residuals from the July retrievals are visible in each case and all data is centred on a latitude of 22.0$\degree$S. Model 1 represents $R$ = 10.0 µm, $V$ = 1.0 radii particles while model 2 is $R$ = 5.0 µm, $V$ = 0.5. Both models have the same refractive index of 1.2 + i0.005. (a) Outside the GRS, the two models are comparable and both perform considerably better than the corresponding NH$_3$ ice model with $R$ = 10.0 µm, $V$ = 5.0 particles. The grey shaded region indicates the data uncertainty. A residual between the data and the model that extends beyond 1 standard deviation is indicated at 10.2 µm. Although the magnitude of this residual is lower inside the GRS, it is believed to be the result of calibration issues. Discrepancies in the 8.5 - 8.8 µm range are most likely the result of wavelength offsets that have not yet been fully corrected. (b) In the centre of the GRS, both models are comparable to the NH$_3$ ice model, with model 2 indicating marginal improvement. (c) and (d), modelled cross-sections of the aerosols, displaying the 9.1 µm and 9.4 µm NH$_3$ ice features that are absent in models 1 and 2.}
            \label{aero_test_fig}
        \end{figure}

        The inability to observe the vertical extent of the multiple aerosol layers in this spectral range \cite{gg, fletcher_2009ph3cirs_paper, fletcher_2010grs_paper} resulted in us modelling the aerosol opacity as a single, compact layer centred at a base pressure (P$_b$). `Aerosol' will hereafter only refer to this single modelled layer encompassing the opacities of all the main cloud decks and the overlying hazes. At altitudes above the P$_b$ pressure level, aerosol concentration decreased with a prior FSH relative to the atmospheric scale height, which was determined at the start of each iteration using the temperature profile via the hydrostatic equation. For altitudes below this pressure level, the concentration dropped exponentially towards zero and therefore we were not able to sound aerosol populations at altitudes below the P$_b$ pressure level. Due to the lack of information on the vertical distribution of aerosols, both $P_b$ and FSH were fixed to values of 1,000 mbar and 0.1 respectively in this retrieval. The lack of spectrally identifiable features from these cloud layers in most of the FOV meant we were also unable to deduce the composition of the observed aerosols. As a result of this, we had to make assumptions about the particle radii and refractive indices. Although NEMESIS is capable of retrieving particle sizes, detailed knowledge of the optical properties of the aerosol layers are required to serve as a prior, otherwise the degeneracy between these variables causes the problem to become ill-constrained. Particle radius was assumed to follow a standard gamma distribution similar to the values used by previous mid-infrared studies \cite{hh, wong_nh3icev2_paper, jj, ddd, fletcher_2009ph3cirs_paper, fletcher_2016texes_paper}. The (unconstrained) radius and variance combination chosen was simply the one that gave the best $\chi^2$ between the model and the spectrum. Radius ($R$) and variance ($V$) combinations ranging from 0.1 µm to 20.0 µm were tested alongside real and imaginary refractive index values of 1.0 - 2.4 and 0.001 - 1.000 respectively. A final refractive index of 1.2 + i0.005 was chosen for this layer. For the particle radii/variance, both $R$ = 10.0 µm, $V$ = 1.0 and $R$ = 5.0 µm, $V$ = 0.5 provided satisfying spectral fits. A comparison of these two, named models 1 and 2 respectively can be seen in Fig. \ref{aero_test_fig}. Model 2 was subsequently chosen for the radii/variance based on this comparison. Stronger constraints on both the altitudes and size distributions of these aerosol layers will be determined through analysis of scattered sunlight in the 4.90 - 7.30 µm range in a forthcoming study. Again, the prior values for this model can be seen in Table \ref{prior_stats}.

        \begin{table}
            \centering
            \begin{tabular}{c c c c}
                \hline
                Species & Parameter & Type & Prior value\\
                \hline
                Temperature & T & Free & Variable Profile\\
                Aerosol 1 & P$_{b}$ & Fixed & 1000 mbar\\
                 & $\tau$ & Free & 1.0 ± 0.1\\
                 & FSH & Fixed & 0.1\\
                 & n & Fixed & 1.2 + i0.005\\
                 & R & Fixed & 5.0 µm\\
                 & V & Fixed & 0.5\\
                Ammonia & P$_{knee}$ & Fixed & 800 mbar\\
                 & P$_{top}$ & Fixed & 100 mbar\\
                 & VMR & Free & 200 ± 20 ppm\\
                 & FSH & Free & 0.15 ± 0.05\\
                Phosphine & P$_{knee}$ & Fixed & 500 mbar\\
                 & P$_{top}$ & Fixed & 100 mbar\\
                 & VMR & Free & 0.5 ± 0.4 ppm\\
                 & FSH & Free & 0.30 ± 0.10\\
                \hline
            \end{tabular}
            \caption{A-priori values and uncertainties used for the variable species in the retrievals. Including; temperature (T), base Pressure (P$_{b}$), integrated opacity ($\tau$), Fractional Scale Height (FSH), aerosol refractive index ($n$), particle radius ($R$), radius variance (assuming a standard gamma distribution) ($V$), knee pressure (P$_{knee}$), top pressure (P$_{top}$) and Volume Mixing Ratio (VMR).}
            \label{prior_stats}
        \end{table}
        
        Although a minimum in para-H$_2$ has previously been observed over the GRS \cite{sada_1996grsvoyager_paper, simon-miller_2002grs_paper, fletcher_2010grs_paper}, contribution of this molecule is primarily at wavelengths longer than 15 µm. Since this is outside the spectral range being modelled in this study, an equilibrium Para-H$_2$ profile that varied with altitude was calculated based on the prior temperature profile \cite{irwin_nemesis_paper}. This profile was constant in both the longitude and latitude directions and was invariant throughout the retrieval process. Hydrocarbons (excluding CH$_4$) do have contribution within the 7.30 - 10.75 µm region, such as the C$_2$H$_2$ feature centred on 7.50 µm, the CH$_3$D feature centred on 8.60 µm and the C$_2$H$_4$ feature centred on 10.50 µm. However these features are small compared to the other contributions at these wavelengths and it is difficult in this data to distinguish these features from the high-flux continuum. In setting up our retrieval architecture, we tested for sensitivity to these species and found that the contribution in this wavelength range was negligible. We also experimented with scaling these abundances during the retrieval, and found no evidence of a need to vary these gases from the prior.  Therefore, for the modelling conducted in this study the abundances of all hydrocarbons were assumed to be spatially uniform. An analysis of the hydrocarbon features beyond 10.75 µm will be the topic of a future paper.
        \par
        Due to the high spectral resolution of the instrument, we adopted a two-stage approach to constrain the full range of altitudes available to us for each species as listed below:
        
        \begin{enumerate}
            \item \textbf{Troposphere fit}\\
                The 8.10 - 10.75 µm (ch1-short - ch2-long) was fitted for all spaxels. Spectral uncertainties were multiplied by 4 in this case. Temperature, aerosols, NH$_3$ and PH$_3$ were all retrieved.  The temperature profiles obtained from this stage were used as the prior temperatures for stage 2.
            \item \textbf{Stratospheric Temperature fit}\\
                The resulting temperature, aerosol and gaseous profiles for each spaxel in stage 1 were then used as the prior for a second retrieval, inferring just temperature in the 7.30 - 8.10 µm (ch1-long - ch2-short) range. Stratospheric temperature was assumed to be exclusively controlled by emission from CH$_4$ and its isotopologues, the abundances of which were assumed to be constant for each spaxel. Spectral uncertainties were multiplied by 8.
        \end{enumerate}

        The CH$_4$ Q-branch (7.66 µm) contained contribution from the highest altitudes available in this wavelength range, as seen in Fig. \ref{contribution_plot}, probing the 1 mbar pressure level and above. However, as shown in Supplementary Figure S5 the provided pipeline uncertainties caused this branch to be poorly fitted. We therefore constrained the spectral uncertainties in the 7.656 µm - 7.678 µm range back down to approximately their native values, reducing by a factor of 10 to force the inversion to fit here.
        \par

        \begin{figure}
            \centering
            \includegraphics[width=\textwidth]{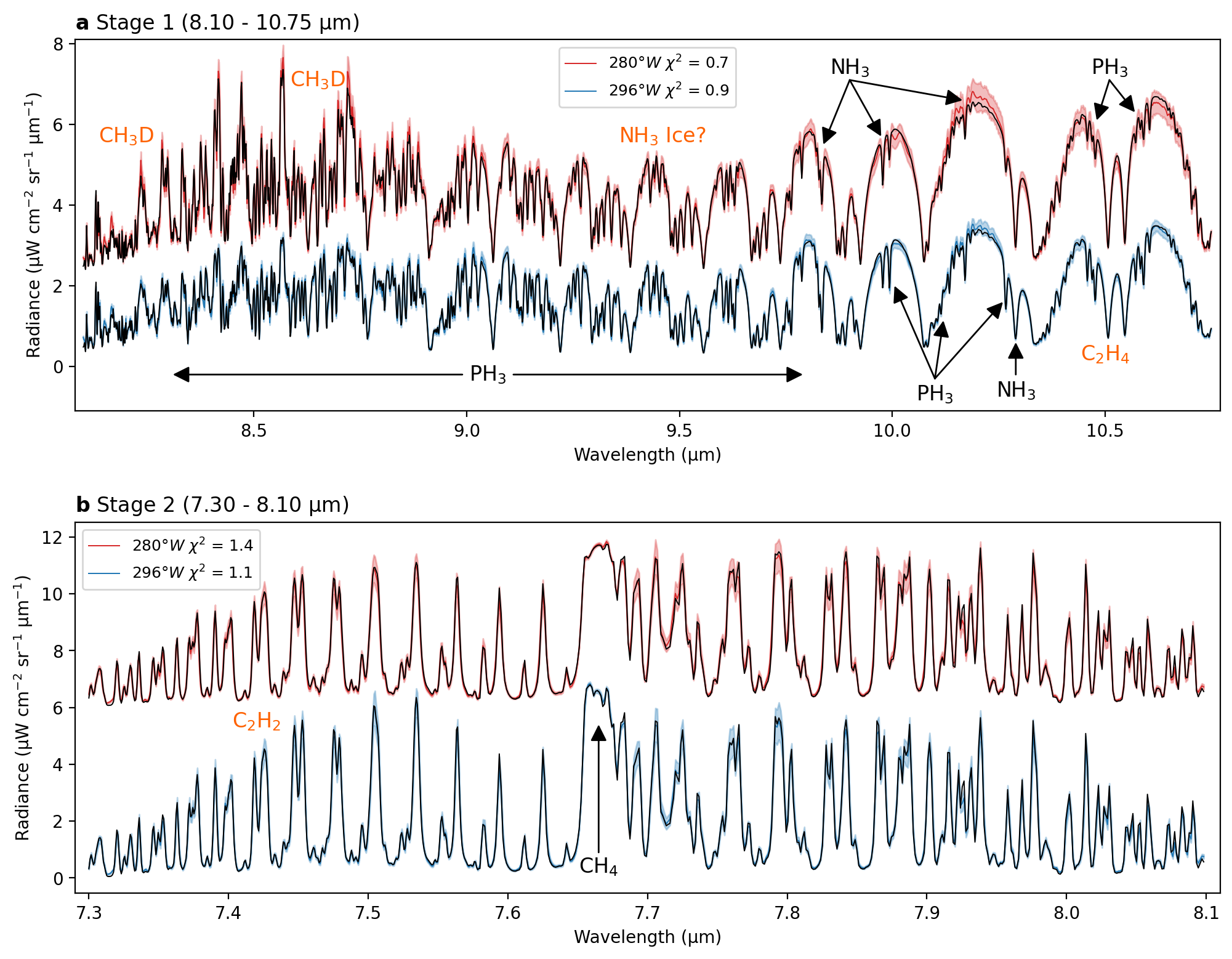}
            \caption{Examples of the quality of the spectral fits (black lines) to the July MIRI/MRS data for inside the GRS (296$\degree$W - blue line) and outside the GRS (280$\degree$W - red line). In both cases the latitude is 21$\degree$S. Major molecular features are denoted in black while minor species are coloured orange. (a) Stage 1 spectral fits sounding tropospheric temperature and species such as aerosols, NH$_3$ and PH$_3$. The residual at 10.2 µm only existed outside the GRS with this region of the spectrum being well fitted inside the vortex. The 280$\degree$W spectra has been offset by +2.0 µW cm$^{-2}$ sr$^{-1}$ µm$^{-1}$ from the 296$\degree$W spectra for clarity. The NH$_3$ ice feature \protect\cite{baines_nh3icev3_paper} and the C$_2$H$_4$ feature are also included, but the high flux of the continuum makes it difficult to observe either of these features.  (b) Stage 2 spectral fits sounding the stratospheric temperature. This spectral region is dominated by the emission features of CH$_4$. The 280$\degree$W spectra has been offset by +6.0 µW cm$^{-2}$ sr$^{-1}$ µm$^{-1}$ from the 296$\degree$W spectra for clarity. In all spectra, retrieval uncertainty is indicated by a shaded region of the same colour as the data.}
            \label{example_fit}
        \end{figure}

        Spectral fits for each of these stages are shown in Fig. \ref{example_fit}. A residual was observed in both this and Fig. \ref{aero_test_fig} at 10.2 µm. Although there are no observed MIRI/MRS instrument calibration artefacts at this wavelength, there are also no known molecular features that we have not already accounted for in the NEMESIS models. This residual is therefore more likely to be an instrumentation artefact, with the absense of this feature inside the vortex being due to the lower brightness of this region. The two stage methodology could not be used for the eastern tile in the August epoch due to only the ch2-short data being available. A separate retrieval; Stage 2 Auxiliary (AUX) for stratospheric temperature was conducted for each tile using just the ch2-short 7.5 - 8.1 µm data to construct a mosaic including the August East data. Aside from aerosols, which were assumed to have a negligible effect on the jovian stratosphere at these wavelengths, the same prior assumptions for Stage 1 were used. Spectral uncertainties were multiplied by 8, with the exception of the Q-branch, the uncertainties of which was again multiplied by 0.8, similar to Stage 2.
        \par
        The uncertainties in the retrieved parameters were hampered by several issues; the tendency of the JWST calibration pipeline to underestimate spectral uncertainties as discussed in Section \ref{data_pipeline}, our choice of profile parameterisation for the gas profiles and degeneracies between each of the 4 parameters such that small variations in the spectra could manifest as different combinations of retrieved parameters. The formal optimal estimation retrieval uncertainties provided by NEMESIS were therefore likely to be an underestimate. For these reasons, although the error bars shown in the graphs in Section \ref{results} display the 1-sigma (1$\sigma$) uncertainties, the contour lines are placed 3$\sigma$ - 5$\sigma$ apart to aid in interpreting variations. We do caution the reader that due to our limited knowledge of the data uncertainty, these are not formal statistical uncertainties and we refrain from overinterpretation of the data, particularly of features that lack multiple sets of observations that are not reproducible between the two epochs.

\section{Retrieval Results}
\label{results}

    \subsection{Stratospheric Temperature Structure}
        \label{results_temp}

        \begin{figure}
            \centering
            \includegraphics[width=\textwidth]{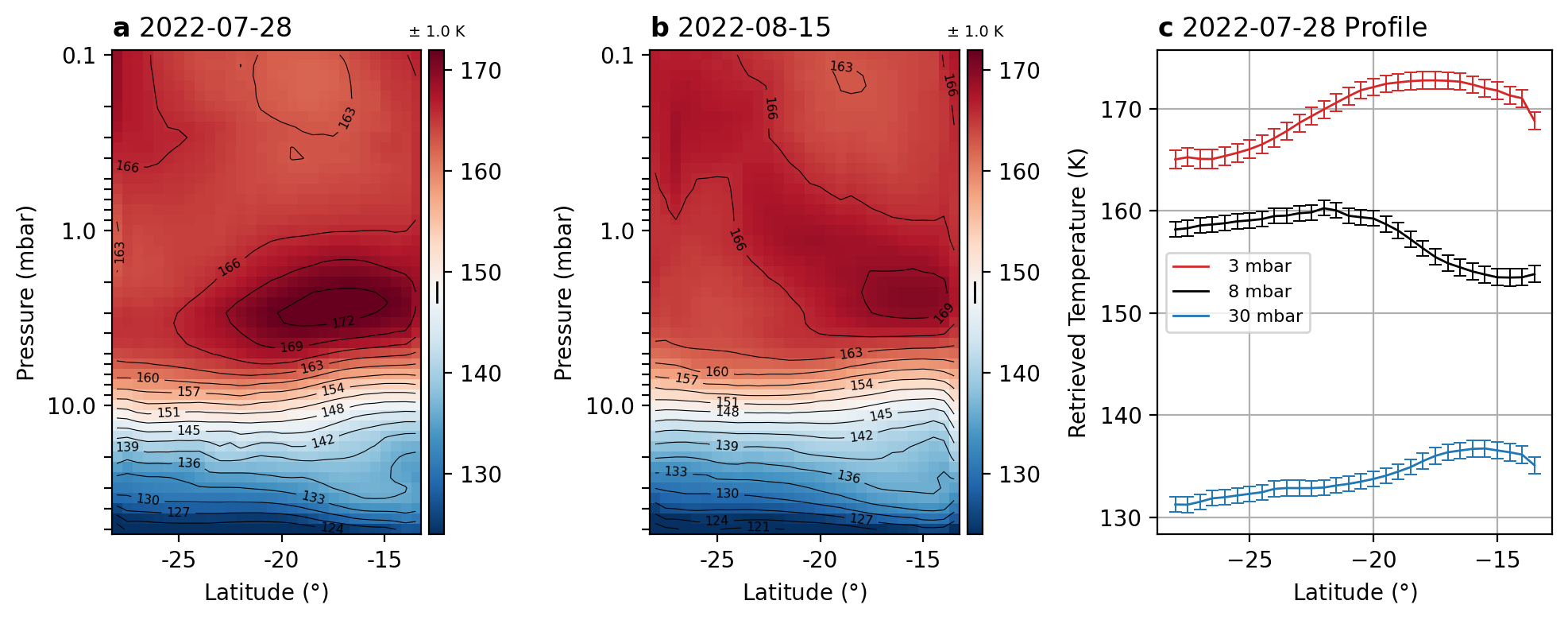}
            \caption{(a) Zonal median of July retrieved stratospheric temperatures over the range of retrieved longitudes from the 7.3 - 8.1 µm spectral range. (b) Similar plot for August. The vertical lines in the colour-bars indicate median retrieved temperature uncertainty. The size of these bars are 1$\sigma$ and the numeric size of this uncertainty is indicated above the colour bars. To aid in interpretation of the data, the contour lines are spaced by 3$\sigma$. (c) Temperature slices extracted from (a) for 3 mbar, 8 mbar and 30 mbar, indicating the differing zonal temperature structure with altitude.}
            \label{stratos_temps1}
        \end{figure}

        \begin{figure}
            \centering
            \includegraphics[width=\textwidth]{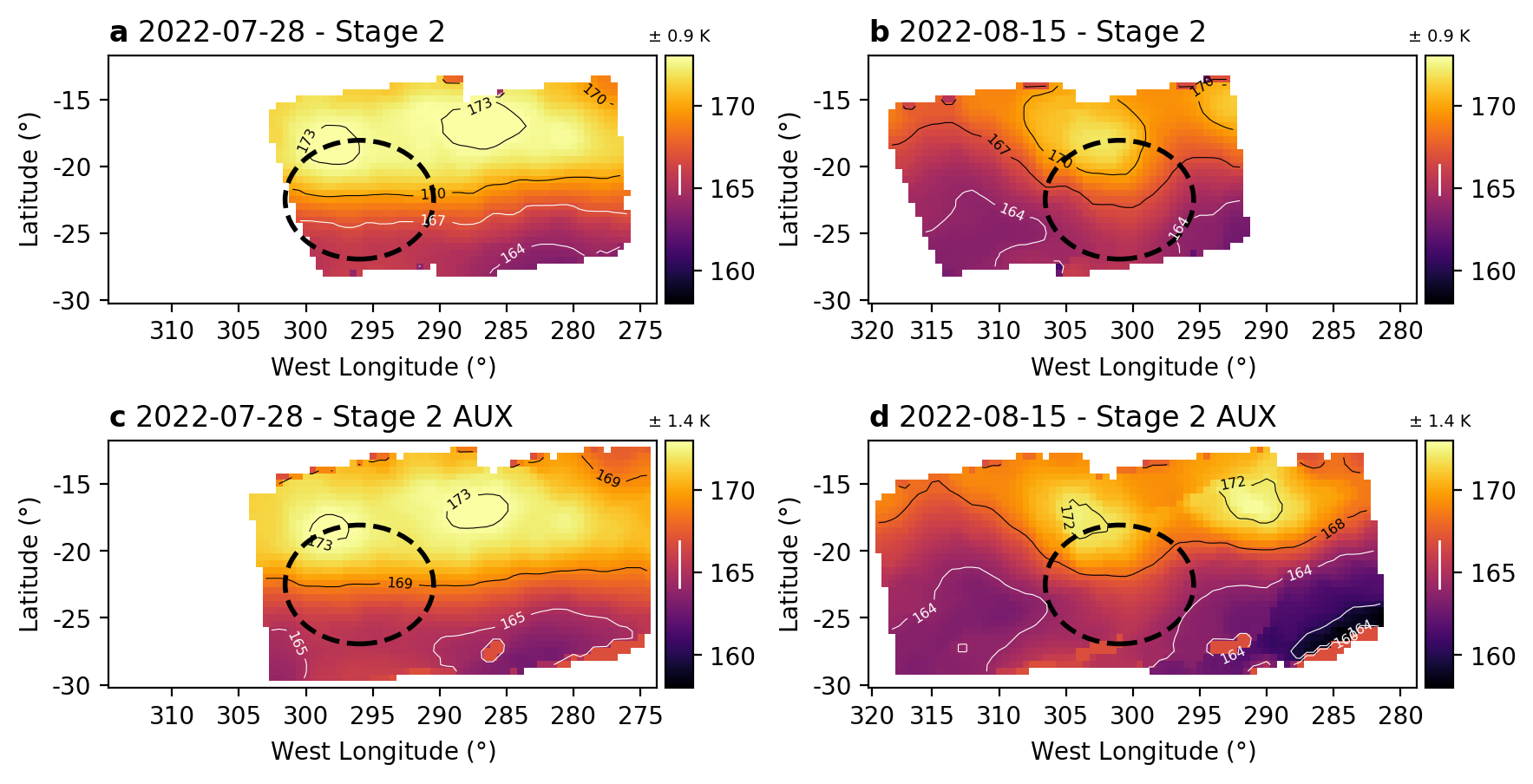}
            \caption{Retrieval of the stratospheric temperature structure at 3 mbar. All FOVs are centred on the longitude of the GRS. The inhomogeneous meridional structure was observed in both the Stage 2 (a and b) and Stage 2 AUX (c and d) retrieval process (the latter simply adding more coverage to the east in August). The smaller FOV of the Stage 2 data is a consequence of using ch1-long data, which has a smaller FOV than ch2-short. Two hot-spots were observed in the upper portion of the FOV to be co-moving with the GRS below, indicated by the dashed black circle in each case. Median 1$\sigma$ uncertainties are indicated by a vertical white line in the colourbar with the numerical uncertainty indicated above the colourbar. The contour lines are spaced by 3$\sigma$ to aid in interpretation of the data.}
            \label{stratos_temps2}
        \end{figure}

        Fig. \ref{stratos_temps1} a and b displays the median stage 2 north-south temperature structure over the range of longitudes retrieved for July and August. Fig. \ref{stratos_temps1}c displays extracted temperature slices from 3 mbar, 8 mbar and 30 mbar where a triple stacked alternating warm and cold temperature anomaly structure (relative to the surroundings) was observed centred on 17$\degree$S. The contrast between these anomalous cells and the surroundings was typically 5-7 K (with a 3$\sigma$ uncertainty of ±3 K) in each case. This structure most likely corresponds to the extra-tropical manifestation of the jovian Quasi-Quadrennial Oscillation (QQO) \cite{ppp, qqq}, also known as the Jupiter Equatorial Stratospheric Oscillation (JESO). The vertical stratospheric temperatures oscillate with altitude above the equator in a pattern of warm and cold cells. Over a four year period these anomalies propagate downwards so the deepest warm temperature anomaly will transition to a cool anomaly and back to a warm anomaly within this time frame. Either side of this are two off-equatorial systems located at ±13$\degree$ latitude that follow the same vertical pattern as at the equator except the temperature anomalies are the opposite to that of the equator.
        \par
        \citeA{56} observed a cold EZ QQO anomaly bounded by two warm anomalies in the horizontal direction at altitudes deeper than 20 mbar in late 2018. Assuming no change to the period of the QQO, the jovian stratosphere would be expected to have returned to the same structure in late 2022 when the MIRI observations were made. This is indeed the case, with the southern 30 mbar warm anomaly located at 13$\degree$S being visible in Fig. \ref{stratos_temps1}c. At 8 mbar the second, higher-altitude QQO cell can be seen on the northern edge of the FOV. As temperature gradients are in thermal wind balance with the zonal winds, this has consequences for the potential variability of the NIRCam-derived winds of \citeA{hueso_2023nircam_paper}.
        \par
        Fig. \ref{stratos_temps2} displays maps comparing the 3 mbar retrieved temperature structure for stage 2 and stage 2 AUX. A further 3D map of this temperature structure in shown in Supplementary Figure S7. Few differences were observed between these two distinct retrieval processes. This altitude displays meridional as well as zonal variation in the form of two hot-spots located in the northern portion of the FOV at (291.0$\degree$W, 17.0$\degree$S) and (304.5$\degree$W, 17.5$\degree$S) respectively in August. These warm anomalies will be referred to as Thermal Mid-atmosphere Anomaly 1 and 2 (TMA 1 and TMA 2) hereafter. First indicated in Fig. \ref{TB_comp}, TMA 1 was observed to be co-moving with the GRS, despite a 5.5$\degree$ drift in longitude between the two epochs as seen in Table \ref{hs_stats}. Temperature differences for each of these features between epochs may be due to differing observing geometry. The temperature anomalies of these two features also become more localised in the 18 days separating observations. This is particularly true for TMA 2 and may be the cause of the apparent drift in position of this feature relative to the GRS. As a result, TMA 2 may also be co-moving with the GRS. A further two hot-spots were observed by VLT/VISIR east of the GRS at similar latitudes to TMA 1 and 2 in Fig. \ref{general_context_figures}c that may also be connected these features.

        \begin{table}
            \centering
            \begin{tabular}{c c c c c c c}
                \hline
                Feature & Epoch & Position ($\degree$) & $\Delta$ lon ($\degree$) & $\Delta$ lat ($\degree$) & T$_{TMA}$ (K) & T$_{mean}$ (K) \\
                \hline
                TMA 1 & July & (298.5, -18.5) & 2.5 & 4.0 & 173.3 ± 1.4 & 169.8 ± 1.4 \\
                 & Aug & (304.5, -17.5) & 3.5 & 5.0 & 172.4 ± 1.4 & 165.9 ± 1.4 \\
                \hline
                TMA 2 & July & (287.5, -17.5) & -8.5 & 5.0 & 173.5 ± 1.4 & 169.6 ± 1.4\\
                 & Aug & (291.0, -17.0) & -10.0 & 5.5 & 172.9 ± 1.3 & 166.1 ± 1.4\\
                 \hline
            \end{tabular}
            \caption{Retrieved stage 2 AUX parameters associated with the 3 mbar hot-spots (TMA 1 and TMA 2 respectively). Stage 2 AUX was used due to TMA 2 not being visible in stage 2. The longitude and latitude positions of these features is provided alongside $\Delta$ lon and $\Delta$ lat relative to the position of the GRS, taken as (296.0$\degree$W, 22.5$\degree$S) and (301.0$\degree$W, 22.5$\degree$S) in July and August respectively. All positions have uncertainties of ±0.5$\degree$ The maximum recorded temperature within the hot-spots ($T_{TMA}$) is compared to the mean temperature across the entire FOV ($T_{mean}$) for that altitude, both with 1$\sigma$ uncertainties.}
            \label{hs_stats}
        \end{table}

        \subsection{Tropospheric Temperature structure}
        \label{tropos_temps_section}

        \begin{figure}
            \centering
            \includegraphics[width=\textwidth]{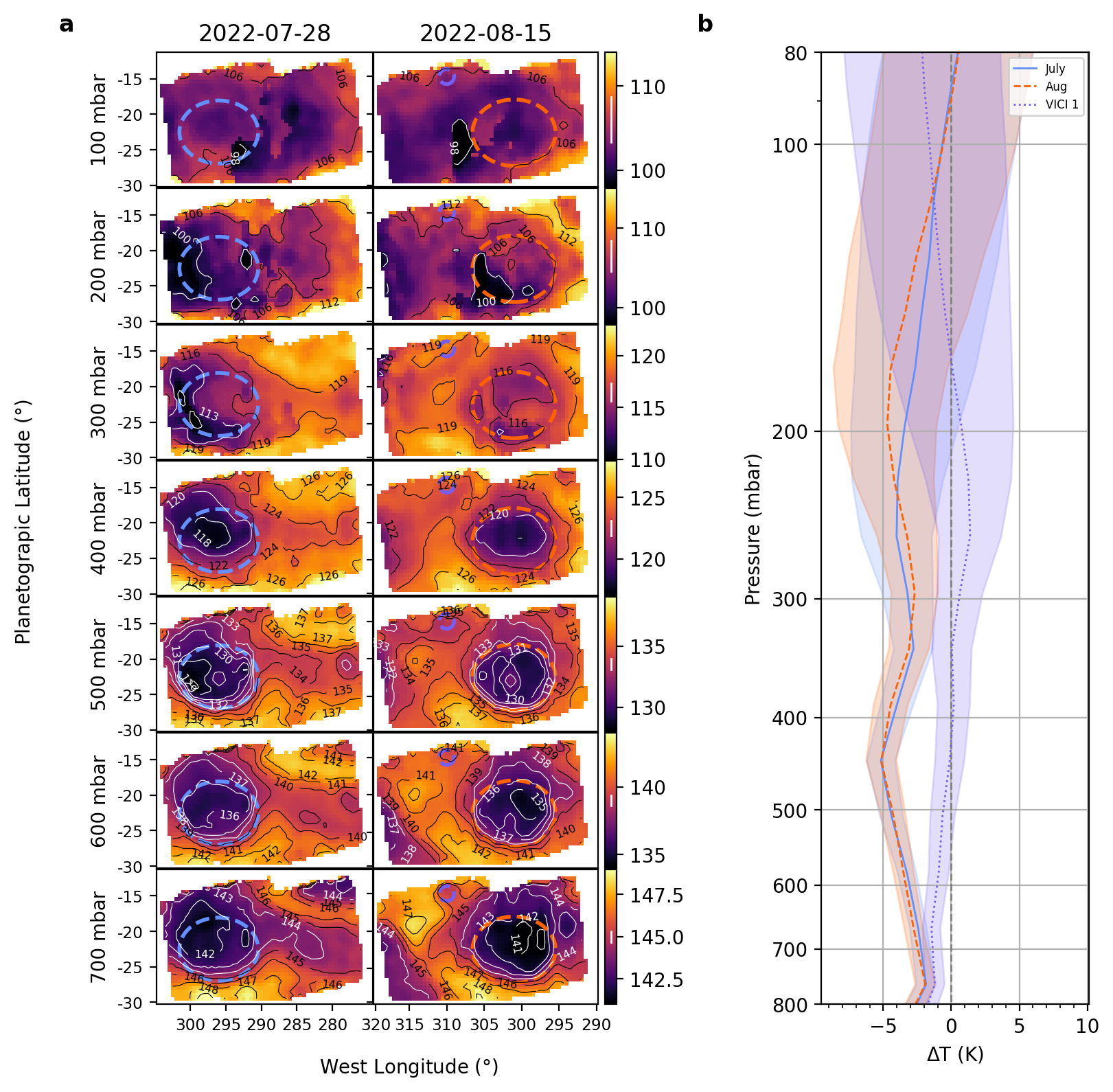}
            \caption{(a) Retrieval of the tropospheric temperatures for altitudes higher than the 800 mbar pressure level (limit of contribution for this wavelength range). The larger circles indicate the ring of peak wind velocities on the edge of the GRS. The blue smaller circles centred on (310$\degree$W, 15$\degree$S) in August correspond to the position of the VICI 1 storm. Median 1$\sigma$ temperature uncertainty for each altitude is indicated by a vertical white line in the colourbar in each case. The contour lines are spaced by 3$\sigma$ to aid interpretation. These vary from ±1 K at 700 mbar up to ±8 K at the tropopause due to a lack of contribution from altitudes higher than the 200 mbar pressure level.  (b) Mean temperature anomalies ($\Delta$T) inside the rings surrounding the GRS in July, August and the ring surrounding VICI 1. The GRS temperature anomalies are relative to a mean temperature profile determined from within a 2$\degree$ wide annulus with an inner radius 1.5$\degree$ greater than the GRS surrounding the vortex. The VICI 1 storm temperature anomaly is relative to a mean temperature profile determined from within a 2$\degree$ wide annulus with an inner radius 2$\degree$ greater than VICI 1 surrounding the instability. 0 K is denoted by a grey, dashed line. The 1$\sigma$ uncertainties are denoted by the shaded regions. A comparison of the difference in temperatures between the two epochs can be seen in Supplementary Figure S8.}
            \label{tropos_temperature}
        \end{figure}

        \begin{figure}
            \centering
            \includegraphics[width=\textwidth]{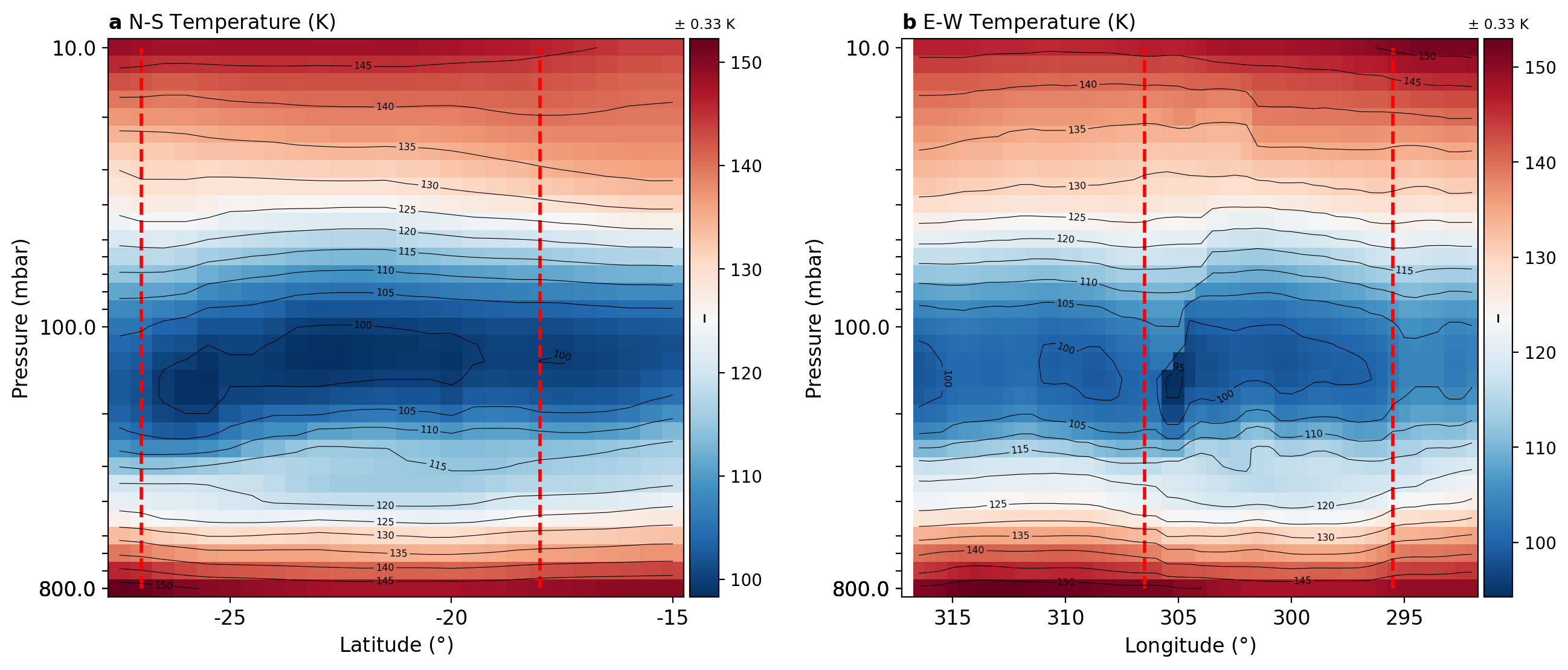}
            \caption{Stage 2 retrieved temperature structure in August for the north-south direction centred on 301.0$\degree$W (a) and the east-west direction centred on 22.5$\degree$S (b). The dashed red lines denote the boundaries of the GRS. Noise within the GRS in the east-west direction is due to the boundary between two MIRI tiles. The median of the 1$\sigma$ retrieved temperature uncertainty for each FOV is visible as a black vertical line within the colourbar and the numerical value of this uncertainty is indicated above the colour bar. 2D maps of the retrieved temperature uncertainty can be seen in Supplementary Figure S9.}
            \label{2D_tempos}
        \end{figure}

        Fig. \ref{tropos_temperature} displays the horizontal variation in tropospheric temperature at a range of altitudes up to the tropopause (a) alongside average temperature profiles inside the GRS and the VICI 1 storm indicating the mean temperature anomaly of both with respect to the surroundings (b). A 3D visual of this temperature structure can also be seen in Supplementary Figure S6. The cold-temperature anomaly of the GRS dominates the FOV of each frame up to the tropopause though, with the greatest temperature anomaly of -4.3 ± 1.8 K (to 3$\sigma$)  with respect to the surroundings being observed at $\sim$400 mbar in both July and August. Although this temperature contrast does decrease at 800 mbar as seen in Fig. \ref{2D_tempos}, this is likely to be an artefact of the retrieval process since there is little contribution below the 800 mbar pressure level. At altitudes higher than the 400 mbar pressure level, this anomaly decreases until the GRS dissipates close to the tropopause (100 mbar). It is difficult to determine what altitude this occurs due to a lack of contribution above 200 mbar causing the temperature structure to become loosely constrained. VICI 1 displays a maximum temperature anomaly of -1.7 ± 1.1 K (to 3$\sigma$) at 800 mbar, typical of an instability that is both smaller and weaker than the GRS. The magnitude of this anomaly then reduces with altitude until it disappears between the 400 - 500 mbar pressure levels. This is consistent with Hubble observations of larger disturbances, which indicated that these plume particles extend to higher altitudes than the surroundings and some can even reach the tropopause \cite{de-pater_2019alma_paper}.
        \par
        Visible at the 500 mbar and 600 mbar pressure levels is a coherent warm feature located at 24$\degree$S close to the centre of the GRS in both epochs. Typically displaying a temperature anomaly of +2.0 ± 1.0 K compared to the surrounding GRS cold temperature anomaly, this feature is visually similar to a warm core observed by \citeA{fletcher_2010grs_paper}. A north-south temperature asymmetry can also be seen in both Fig. \ref{tropos_temperature} and \ref{2D_tempos} at altitudes above the 800 mbar pressure level. Outside the northern boundary of the ring of peak wind velocities (a region corresponding to the white GRS collar), temperatures are consistently 2-3 K (±1 K) cooler than those at the southern collar. Such inhomogeneous structure has been previously observed at mid-infrared wavelengths \cite{flasar_1981grsvoyager_paper, sada_1996grsvoyager_paper, eee, m, simon-miller_2002grs_paper, fff, fletcher_2010grs_paper} and has been tentatively assessed to be caused by the GRS tilting in the north-south direction. North-south asymmetries are also noted in the aerosol, NH$_3$ and PH$_3$ distributions in Sections \ref{aerosol}, \ref{ammonia} and \ref{phosphine} and will be discussed in Section \ref{asymmetries}.

        \begin{figure}
            \centering
            \includegraphics[width=\textwidth]{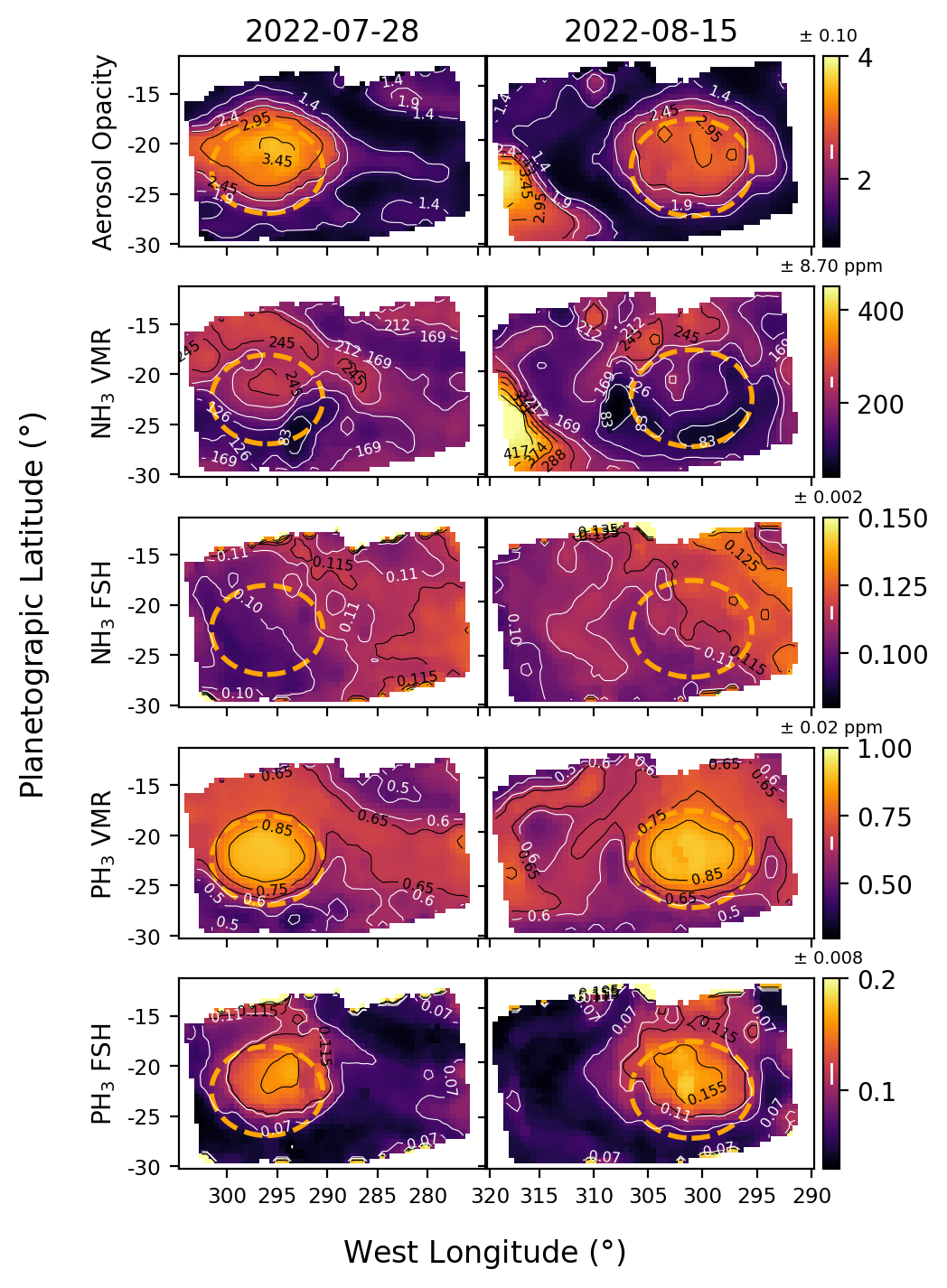}
            \caption{Retrieval of atmospheric variables: integrated aerosol opacity down to 1.0 bar, 800 mbar NH$_3$ VMR and FSH and 500 mbar PH$_3$ VMR and FSH. Both MIRI epochs are displayed for comparison. The position of maximum wind velocities in the outer rim of the GRS is indicated by the dashed orange circle in each case. The median 1$\sigma$ retrieved uncertainty for each of these parameters is presented as a vertical white line within each of the colourbars and the numerical magnitude of this uncertainty is given above the colour bar. The contour lines are spaced by 5$\sigma$ to aid interpretation of the data.}
            \label{nh3_ph3_aero_results}
        \end{figure}

        \subsection{Aerosol distribution}
            \label{aerosol}
            Fig. \ref{nh3_ph3_aero_results} displays the integrated aerosol opacity down to 1.0 bar (dimensionless) within the jovian troposphere for the GRS and surrounding atmosphere. Little change is seen in the 18 days separating observations despite the 4 day rotation period of the GRS. The aerosol distribution coincident with the region of red chromophore was also uniform, implying a stable and well-mixed band of aerosol layers residing above the GRS. Indeed one of the reasons why this material may be darker in colour than the surrounding vortex as seen in Fig. \ref{jup_rgb_troposphere}a and b is because of this region being unusually stable and allowing the aerosol material and the chromophore to be concentrated here \cite{baines_2019grschromophore_paper, sanchez-lavega_2021grsinteractions_paper}. Interestingly, the aerosols within the STrZ, the SEB, VICI 1, the low-opacity GRS collar and the high-opacity GRS vortex are indistinguishable in the mid-infrared, despite the aerosols in question being known to be physically distinct from one another \cite{dahl_2021grscolour_paper}. Within the GRS collar, the northern integrated opacity was almost a factor of 2 greater than the opacity in the south. Both NH$_3$ and PH$_3$ also display north-south asymmetries outside the vortex thought to be related to the vortex tilting in the north-south direction. This is further discussed in Section \ref{asymmetries}.
            \par
            Observations of this FOV in the 890 nm spectral region shown in Fig. \ref{general_context_figures}b indicate that the aerosols within both the GRS and VICI 1 reach higher altitudes than the surroundings \cite{sanchez-lavega_2018grsjunocam_paper, anguiano-arteaga_2021grsaerosol_paper, anguiano-arteaga_2023grsaerosol_paper}. In addition to this, the HST and ground-based context images in Fig. \ref{jup_rgb_troposphere}a and b show that the STrZ and VICI 1 are coincident with a region of high-albedo white clouds while the SEB and GRS aerosols have blue-absorbing properties, suggesting compositional differences between these regions. Differences between these four regions will likely begin to emerge when we model MIRI data shortward of 7 µm. Despite being compositionally indistinguishable, the GRS and STrZ are still separated by a lane of low opacity. A similar lane of low opacity separates the GRS from the SEB in the north-east, with no indication of the GRS hollow being present \cite{rogers_1995jupiter_book}. The visual HST and ground-based observations in Fig. \ref{jup_rgb_troposphere}a and b show that these low opacity lanes coincide with darker regions that surround the GRS. Previous observations of the GRS in the 5 µm region of the spectrum, sensitive to deep tropospheric emission between 2-4 bar indicate a patchy ring of bright emission within this lane, not unlike the 5 µm rings often seen around jovian anticyclones \cite{de-pater_2010vortexrings_paper}. Such depletion of aerosol opacity and emission at 5 µm is suggestive of atmospheric subsidence.

        \subsection{Ammonia distribution}
            \label{ammonia}
            Fig. \ref{nh3_ph3_aero_results} displays both the 800 mbar VMR and the FSH for NH$_3$. Across the FOV, the mean retrieved 800 mbar VMR was 182 ± 50 ppm and 174 ± 43 ppm for July and August respectively (to 5$\sigma$). This is consistent with a 1-2 bar zonal mean abundance in the jovian southern hemisphere, determined using radio observations made by the Karl G. Jansky Very Large Array (VLA) \cite{de-pater_2016vla_paper, depater_2019vla_paper}. Previous observations of the NH$_3$ abundance within the GRS have been inconsistent, with some indicating excess over the GRS \cite{sada_1996grsvoyager_paper} and others indicating depletion \cite{bbb}. No large-scale excess was observed inside the ring of peak vortex velocities compared to the surroundings, similar to the NH$_3$ content inferred by Juno/JIRAM \cite{grassi_2021jiram_paper}. However, a north-south asymmetry was observed outside the GRS, with an arc of NH$_3$ depletion within the southern GRS collar being contrasted by a band of enhancement in the northern collar. This distribution was also observed in 2000 by Cassini/CIRS and in 2014 by IRTF/TEXES \cite{fletcher_2016texes_paper}. The arc of depletion, separating the GRS and the STrZ is coincident with the dark lane of presumed subsidence seen in the aerosol data in Section \ref{aerosol} in the south. The variations of VMR inside the GRS vortex are most likely an artefact of the retrieval process due to the problems inferring the retrieval uncertainties, discussed in Section \ref{NEMESIS}. However, the large-scale north-south asymmetry present in both epochs, coinciding with north-south asymmetries in the temperature structure, the aerosol opacity distribution and the PH$_3$ VMR and FSH maps are consistent with previous observations. The largest excess, exceeding 400 ± 43 ppm is within the STrZ in the lower left portion of the August FOV. This does concur with previous near \cite{grassi_2021jiram_paper} and mid-infrared observations \cite{fletcher_2016texes_paper} that suggest that zones display excess NH$_3$ abundance and belts display depletion. Despite the observed VMR excess in the STrZ however, the FSH of this region is comparable to the mean of 0.11 ± 0.01. One possible removal process for excess high-altitude NH$_3$ here could be the generation of the high-albedo banks of cloud that dominate the visible appearance in this region. A similar localised excess of 280 ± 43 ppm was observed within VICI 1, potentially the result of vigorous convection driving the NH$_3$ up to higher altitudes from the deep troposphere. Despite this, the FSH within this region is also comparable to the mean of the surroundings.

        \subsection{Phosphine distribution}
            \label{phosphine}
            The historical distribution of PH$_3$ above the GRS is complex and possibly varies in response to global upheaval events such as SEB fades \cite{17}. The mean 500 mbar VMR observed across the FOV, as seen in Fig. \ref{nh3_ph3_aero_results} was 0.65 ± 0.10 ppm (to 5$\sigma$) and 0.66 ± 0.10 ppm in July and August respectively. This is consistent with deep abundances of 0.6 - 0.7 ppm inferred by both IRTF/TEXES \cite{fletcher_2016texes_paper} and Juno/JIRAM \cite{grassi_2020jiram_paper}. We found an excess of 0.90 ± 0.10 ppm centred on the GRS at 500 mbar in both epochs, with a corresponding excess in FSH of 0.17 ± 0.04 (compared to the mean of 0.09 ± 0.04 and 0.08 ± 0.04 for July and August respectively) also suggesting that this PH$_3$ persists to higher altitudes than the surroundings. The distributions of both PH$_3$ and NH$_3$ with respect to the saturation vapour pressure and the aerosol distribution is further analysed in Supplementary Figure S10. The persistence of PH$_3$ to higher altitudes may allow upper tropospheric photochemical reactions involving this molecule that generate the currently unidentified red chromophore \cite{irwin_book_latestedition}, a compelling argument since the ring of peak vortex velocities entirely encase the red chromophore material and the PH$_3$ VMR excess observed within the GRS. A north-south asymmetry similar to the aerosol opacity and NH$_3$ VMR distributions was also observed in both the PH$_3$ VMR and FSH outside the GRS, with the VMR north of the GRS being 40\% greater than south of the vortex and the FSH in the north being twice that of the south. Both these factors concur with the suggestion of upwelling occurring within the northern portion of the white GRS collar and subsidence in the south, this is further discussed in Section \ref{asymmetries}. Within the STrZ, seen in the lower-left corner of the August FOV, an excess of 0.76 ± 0.10 ppm was observed. This is consistent with IRTF/TEXES observations, which indicated excess PH$_3$ within jovian zones \cite{fletcher_2016texes_paper}. The observed excess within the STrZ was also uneven, suggestive of localised regions of convection rather than one large uniform region of upwelling air. VICI 1 displayed an excess of 0.73 ± 0.10 ppm in VMR, consistent with the idea of vigorous upwelling air but with a FSH of 0.11 ± 0.04 that was similar to the FOV mean of 0.09 ± 0.04. This small instability most likely has not existed for sufficient time to generate a high-altitude PH$_3$ excess similar to the much older GRS.

\section{Discussion}
\label{discussion}

    \subsection{Temperature, gaseous and aerosol degeneracies}
        \label{degeneracy}

        \begin{figure}
            \centering
            \includegraphics[width=\textwidth]{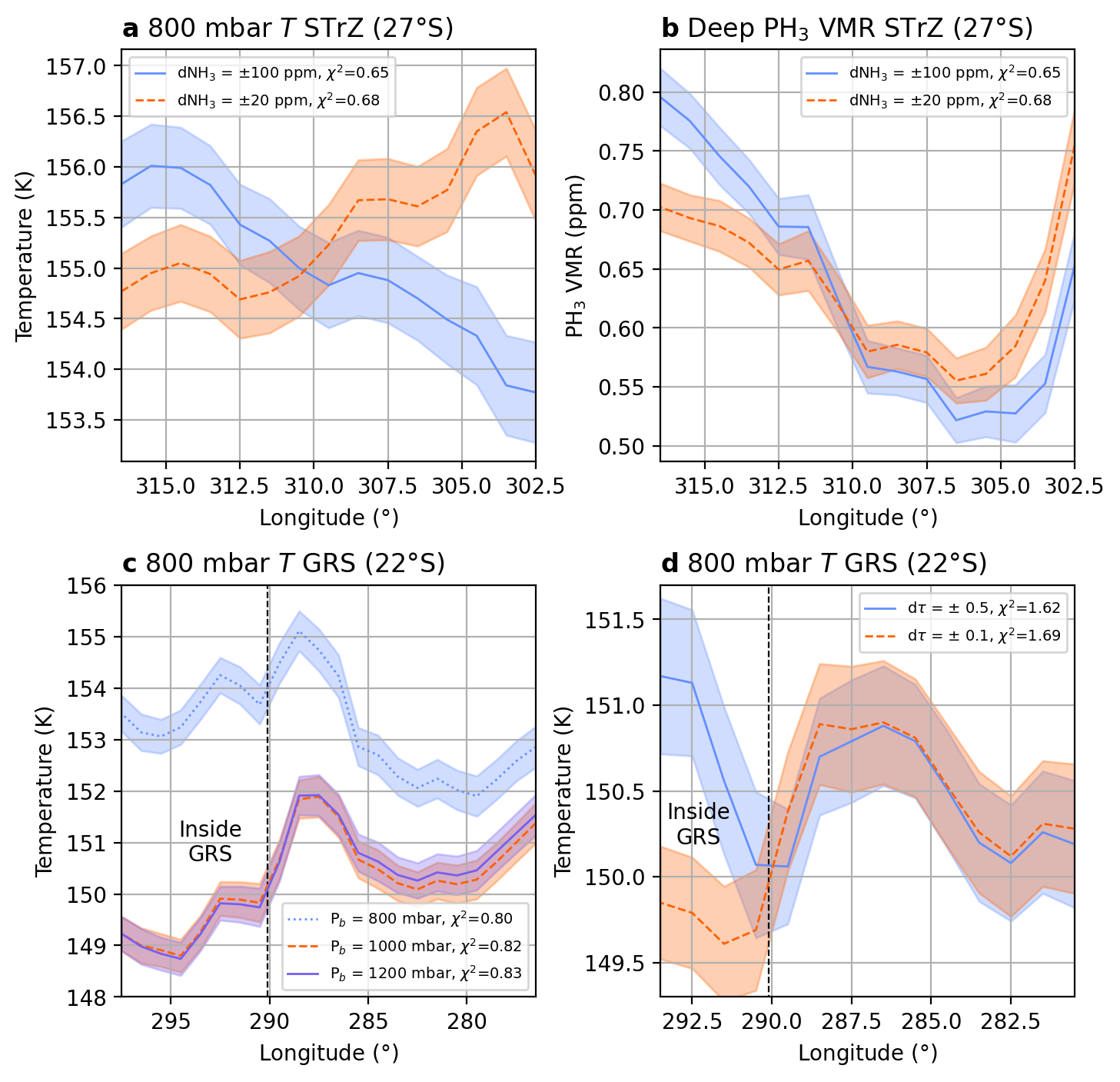}
            \caption{Retrieval of Temperature, NH$_3$, PH$_3$ and aerosols in July for altitudes deeper than the 800-mbar pressure level. For each test, $\chi^2$ is indicated for the westernmost position. Similar $\chi^2$ values within each of these plots hid the variability inherent in these four parameters due to degeneracies between them. (a) The temperature ($T$) structure of the GRS wake was found to be highly dependent on the chosen prior uncertainty for NH$_3$ VMR (dNH$_3$). Higher values for this uncertainty could also cause the retrieved NH$_3$ VMR within the STrZ to exceed 1000 ppm, entirely inconsistent with theory. A smaller value of ±20 ppm for dNH$_3$ was therefore chosen. (b) Similar plot within the STrZ for retrieved PH$_3$ VMR, indicating that changes to dNH$_3$ affected the resulting PH$_3$ profile too, although the effect here was less severe than for temperature. (c) Retrieved temperature within the GRS vortex, demonstrating the effect of changing the aerosol layer base pressure (P$_b$). No trend could be discerned between these two parameters, with P$_b$ eventually being chosen such that the GRS remained cold at depth. (d) A degeneracy was also observed between the GRS temperature at depth and aerosol opacity ($\tau$). A less constrained $\tau$ allowed the GRS to be warm at depth, inconsistent with modelling.}
            \label{degeneracy_fig}
        \end{figure}
        
        One of the primary objectives of this study was to use JWST/MIRI's enhanced sensitivity, spectral resolution and extended wavelength range to separate the contributions of temperature, NH$_3$, PH$_3$ and aerosol abundance in this wavelength range for the first time. JWST/MIRI provided good sensitivity to the distribution of the parameters retrieved at altitudes above the 800 mbar pressure level. However, degeneracies between parameters at deeper pressures near 800 mbar proved to be more challenging and represent the limit of what we can do with spectra in this wavelength range. Initial tests retrieving these species quickly revealed a degeneracy between knee pressure and retrieved VMR for the gaseous species, with higher altitude knee pressures consistently resulting in a lower retrieved VMR as seen in Section \ref{NEMESIS}. The inability to break this degeneracy forced us to alter the knee pressures such that the retrieved NH$_3$ VMRs were consistent with Juno/MWR observations \cite{li_2017mwrnh3_paper, moeckel_2023mwr_paper} and the retrieved PH$_3$ VMR were consistent with Cassini/CIRS observations \cite{fletcher_2009ph3cirs_paper}. Further degeneracies between temperature, NH$_3$ and PH$_3$ VMR were discovered within the STrZ, as seen in Fig. \ref{degeneracy_fig}a and b. The loosely constrained initial prior NH$_3$ VMR allowed the retrieved values to exceed 1000 ppm, entirely inconsistent with previous studies \cite{depater_2019vla_paper, grassi_2021jiram_paper, 51, bbb, l, ddd, li_2017mwrnh3_paper, moeckel_2023mwr_paper, fletcher_2010grs_paper, fletcher_2016texes_paper}. However, constraining this species also altered both the retrieved temperature and PH$_3$ profiles. Furthermore, aerosol base pressure ($P_b$) was observed to interfere with the retrieved deep temperature profile, particularly inside the GRS as seen in Fig. \ref{degeneracy_fig}c. Modelling predicts that the GRS should be cold down to the mid-plane \cite{de-pater_2010vortexrings_paper, palotai_2014anticyclonemodelling_paper, 60} and in the absense of contribution below 800 mbar, we expect the retrieved temperature structure to continue the cooling trend of higher altitudes. Certain aerosol altitude/temperature combinations actually caused the GRS to be uniformly warm at depth compared to the surroundings even though no indication of a warm GRS is observed at any point in the spectral data. This issue was particularly pronounced in the July eastern tile and the August central tile, both of which were taken closer to the jovian limb as seen in the emission angles of Table \ref{observation_table}. Finally, a further degeneracy was observed between temperature and aerosol opacity inside the GRS vortex, a region known to be elevated in opacity compared to the surroundings. For the observations that were taken close to the jovian limb such as July/East and August/Centre and prior to constraining the prior uncertainties, the retrieved aerosol opacities inside the GRS vortex were an order of magnitude higher than their disc-centre counterparts and the corresponding vortex temperature was warmer than the surroundings as seen in Fig. \ref{degeneracy_fig}d.
        \par
        In much the same way as the gaseous knee pressure degeneracies were addressed, P$_b$, 800 mbar NH$_3$ VMR uncertainty, 1-10 bar temperature uncertainty and the prior uncertainty in integrated opacity had to be adjusted until a combination was found that provided NH$_3$ and temperature results that were consistent with previous observations and gave a goodness-of-fit close to 1. There were no independent ways of determining temperatures below 800 mbar for this study and therefore the retrieval instead relied on a single prior temperature profile. This consisted of Galileo probe measurements for deep tropospheric temperatures below 1 bar, which were measured in a highly anomalous North Equatorial Belt (NEB) hot-spot. Although the retrievals at shallower pressures were well-constrained, the lack of a temperature structure determined independently from this MIRI data resulted in the distribution of the deep component of the retrieved parameters still being determined based on our choice of priors.

    \subsection{GRS Vortex Structure}
            \label{asymmetries}
    
            Fig. \ref{nh3_ph3_aero_results} indicates enhanced aerosol opacity, NH$_3$ VMR, PH$_3$ VMR and PH$_3$ FSH  abundances outside the northern limb of peak wind velocities, corresponding to the white GRS collar. All this concurs with previous observations indicative of the GRS tilting, with the GRS collar flowing deeper on the southern limb of the storm \cite{m, sada_1996grsvoyager_paper, eee, simon-miller_2002grs_paper, fff, fletcher_2010grs_paper, fletcher_2016texes_paper}. This implies enhanced upwelling occurring north of the vortex, concurrent with the observation of colder deep temperatures generated through adiabatic cooling and enhancements in aerosol densities, NH$_3$ and PH$_3$ VMR. A corresponding subsiding arc occurs in the south of the ring of peak vortex winds, inducing warmer temperatures and depletion of aerosol integrated opacity, NH$_3$ and PH$_3$ VMR. The effects of this tilt on the temperature structure are observed at a range of tropospheric pressures, implying this may be a deep, rather than superficial feature of the GRS. This is concurrent with the results for the aerosol study performed by \citeA{anguiano-arteaga_2023grsaerosol_paper}. Results from Juno/MWR also suggest that the NH$_3$ dynamical processes within the GRS may operate deeper than the 240 mbar pressure level \cite{bolton_2021mwr_paper}, suggesting that the NH$_3$ north-south asymmetries may persist at a range of altitudes as well.
            \par
            Within the vortex, the tropospheric temperatures were not uniformly cold, but instead had small-scale (\textless 500 km) variability that changed over the 18 days separating observations. The degeneracies outlined in Section \ref{degeneracy} make it difficult to determine whether these are the manifestation of a thermal or NH$_3$ effect. The 24$\degree$S thermal anomaly within the GRS is one such feature. This coincides with the region of darkest chromophore material observed inside the vortex in the visible-wavelength data. It is possible that this thermal anomaly could be related to the warm-core feature observed by \citeA{fletcher_2010grs_paper}. Alternatively, it could be an internal localised patch of differing temperature/NH$_3$ abundances related to gyres within the GRS similar to those observed by JunoCam \cite{sanchez-lavega_2018grsjunocam_paper}. The formation and lifetime of such features is still poorly understood \cite{pppp}. As such, it is difficult to determine if the features observed in July and August are the same internal structures or not.
            \par
            The cold temperature anomaly of the GRS was also observed to be displaced 2$\degree$N compared to the ring of peak wind velocities, tending to align with the aerosol opacity rather than the velocity maps. If the uniform distribution of NH$_3$ FSH in Fig. \ref{nh3_ph3_aero_results} is due to the excess NH$_3$ in the northern collar condensing into the excess aerosol opacity observed in this location, then the lower temperatures in the northern collar caused by the off-centre temperature anomaly could be the instigator of this condensation process. Conversely, PH$_3$ was observed to be in excess within the GRS ring of peak wind velocities. This is most likely due to the different dissipation mechanisms these molecules undergo. Although the anticyclonic nature of the vortex may concentrate this molecule, at tropospheric altitudes it would normally then be photolysed and removed by UV light \cite{de-pater_2010vortexrings_paper}. However, in this case the high-altitude aerosol layers within this region may shield PH$_3$ from photolysis, allowing it to accumulate into the excess we see centred on the GRS in Fig. \ref{nh3_ph3_aero_results}.

    \subsection{Inhomogeneous Stratospheric Structure}
        Although the QQO induces considerable north-south temperature anomalies in the jovian stratosphere, it produces little east-west variation. Consequently, the Hot-Spots observed at 3 mbar are likely to be produced by a different mechanism. The GRS is a high-pressure anticyclone that impinges on the stably-stratified tropopause and lower stratosphere, therefore air flowing east or west will encounter this obstacle. It is possible that this could generate a range of atmospheric waves \cite{conrath_1981grswaves_paper, vvvv, wwww}. Orographic Gravity Waves are one possibility however, the small scale of these features would make them difficult to detect with the spatial resolution of JWST/MIRI unless multiple sources combined to form an unusually large wave packet. Alternatively these hot-spots could be the result of a vertically propagating Rossby Wave, not unlike the Tropospheric Equatorial Rossby Wave responsible for much of the spatial structure in the NEB \cite{6}. The morphology of a Rossby Wave is determined by the Coriolis forces acting on it, resulting in large meridional variation but little zonal structure. Potentially the two hot-spots captured by VLT/VISIR on 2022-08-08 displaced 65$\degree$W and 90$\degree$W of the GRS could be a part of this system too. However, this is difficult to prove with one set of observations for this region. It is also possible that these could be the product of an unrelated Stratospheric wave or circulatory system that happened to co-move with the GRS during the campaign of MIRI observations.
        \par
        Two similar temperature anomalies were observed by VLT/VISIR in 2018 \cite{bardet_2024visirjupiter_paper} and by Voyager/IRIS in 1979 \cite{flasar_1981grsvoyager_paper}. Interestingly, neither Cassini/CIRS nor IRTF/TEXES observed these features in 2000 and 2014 respectively \cite{fletcher_2016texes_paper}. It is therefore possible that these features are sporadic but when they do manifest, they appear at the same longitudes relative to the GRS. \citeA{flasar_1981grsvoyager_paper} commented that a vertically propagating wave disturbance would only produce features downstream of the GRS, making the presence of TMA 1 difficult to explain. Furthermore, a model incorporating a vertically propagating Rossby wave was unable to even account for TMA 2. Ultimately, the propagation of atmospheric waves in the mid-atmosphere in response to a tropospheric disturbance is poorly understood and three observations of these features is insufficient to determine if they are periodic. Modelling of each of the proposed wave mechanisms should be undertaken to better understand the altitudes that these phenomena would manifest as well as their magnitudes. This should be done in conjunction with a series of longer-term observations of the hot-spots to be reasonably confident as to the identity of these features.

    \subsection{Thermal wind analysis}
        \label{thermal_winds}

        The outer-annulus primary circulation of the GRS is known to rotate anticlockwise through the collar. This occurs around a high-pressure anomaly at velocities exceeding 100 m s$^{-1}$. Although centrifugal force on the region of high vorticity at the edge of the GRS invalidates the assumption of geostropic equilibrium, it is still possible  to assume that on average the GRS experiences geostropic balance \cite{1}. Therefore, the horizontal temperature gradients, presented in Fig. \ref{2D_tempos} can be related to the vertical wind shear through  the dry adiabatic thermal wind-shear equations \cite{ll}:

        \begin{equation}
            \label{wind1}
            \frac{\partial u}{\partial z} = -\frac{g}{f T} \left(\frac{\partial T}{\partial y}\right)_P
        \end{equation}

        \begin{equation}
            \label{wind2}
            \frac{\partial v}{\partial z} = \frac{g}{f T} \left(\frac{\partial T}{\partial x}\right)_P
        \end{equation}

        where $x$, $y$ are the east-west and north-south distances in km and $z$ is the vertical altitude above the 1 bar pressure level in km, $u$ and $v$ are the zonal and meridional wind in m s$^{-1}$ respectively, $g$ is the jovian acceleration due to gravity (Calculated values of which can be seen in Supplementary Figure S11), $f = 2 \Omega \sin{\phi_g}$, where $\Omega$ is the angular velocity due to the jovian planetary rotation period of 9.9 hrs, $\phi_g$ is the planetographic latitude, $T$ is the 3D temperature field and $P$ is the atmospheric pressure.
        \par
\begin{figure}
            \centering
            \includegraphics[width=1.0\textwidth]{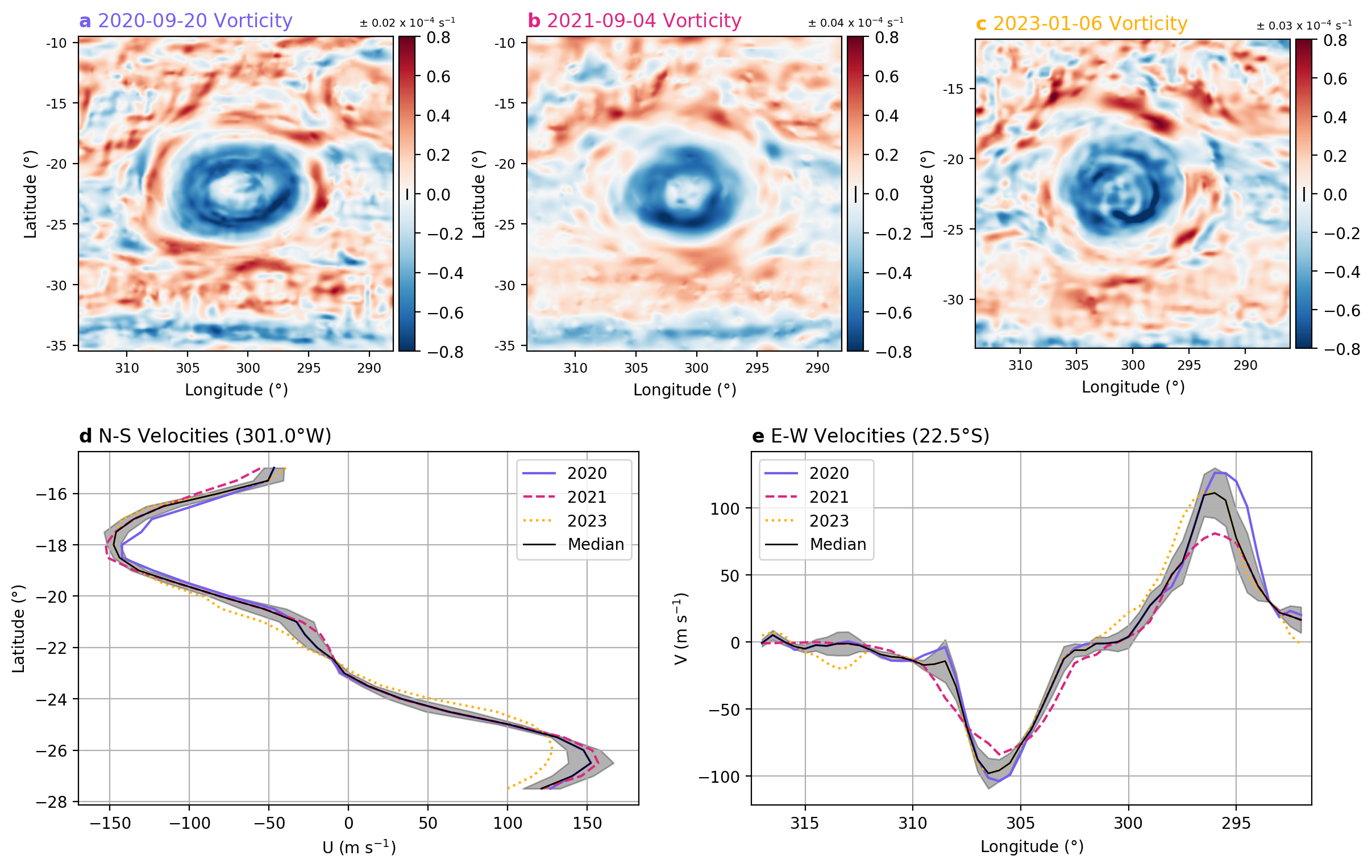}
            \caption{(a), (b) and (c) 2020, 2021 and 2023 vorticity measured from Hubble observations. For the purposes of median combining the wind data, the longitudes were shifted to centre the GRS on 301.0$\degree$W, corresponding to the longitude of the GRS on the date of the August MIRI observations. White corresponds to 0 s$^{-1}$. The uncertainties for these vorticity measurements were derived from a combination of the velocity uncertainties and the uncertainty in the GRS core size. These values are indicated by a vertical black line in the colourbar, with the numerical uncertainty presented above the colourbar. (d) and (e) Derived north-south and east-west winds passing through the GRS centre at (301$\degree$W, 22.5$\degree$S) from the Hubble data for the three epochs, alongside the median of these epochs. This median served as the prior wind for the thermal wind analysis.}
            \label{prior_vorticity_fig}
        \end{figure}

        Equations \ref{wind1} and \ref{wind2} enabled us to integrate the wind velocities within the GRS as a function of pressure. A boundary condition was imposed on this data, consisting of a median of Hubble wind measurements introduced in Section \ref{data_hubble} taken on; 2020-09-20, 2021-09-04 and 2023-01-06. Any shrinkage of the GRS in this time was assumed to be negligible \cite{7}. The measured vorticities for each epoch can be seen in Fig. \ref{prior_vorticity_fig}a-c while the prior wind field used for the thermal wind analysis can be seen in sub-figures d and e. The differences between the Hubble wind measurements were most likely the result of variations in data quality, emission angle and different time intervals between the 4 observations for each epoch.
         
            The altitude of this Hubble prior wind field was initially assumed to be loosely constrained within the 600 - 500 mbar range \cite{aaaaa, west_2004jupitercloud_book}. A lack of radiative transfer modelling of the reflective clouds observed in the visible spectrum is the cause of this poor constraint. A similar issue is also experienced by the NIRCam/F212N filter, loosely constrained to the $\sim$240 mbar range \cite{ddddd, ccccc, hueso_2023nircam_paper}. In addition, the different optical and chemical properties of the GRS may result in the altitude of peak contribution for these 2 wavelength ranges being different inside the vortex compared to both the GRS collar and the wider jovian troposphere. For the purposes of this investigation we assumed that the Hubble wind field was probing the same altitude in both the core and the periphery of the vortex. However, this question is currently difficult to answer without more certainty over the altitudes where peak contribution occurs in the visible and near-infrared range.
            \par

            \begin{figure}
                \centering
                \includegraphics[width=1.0\textwidth]{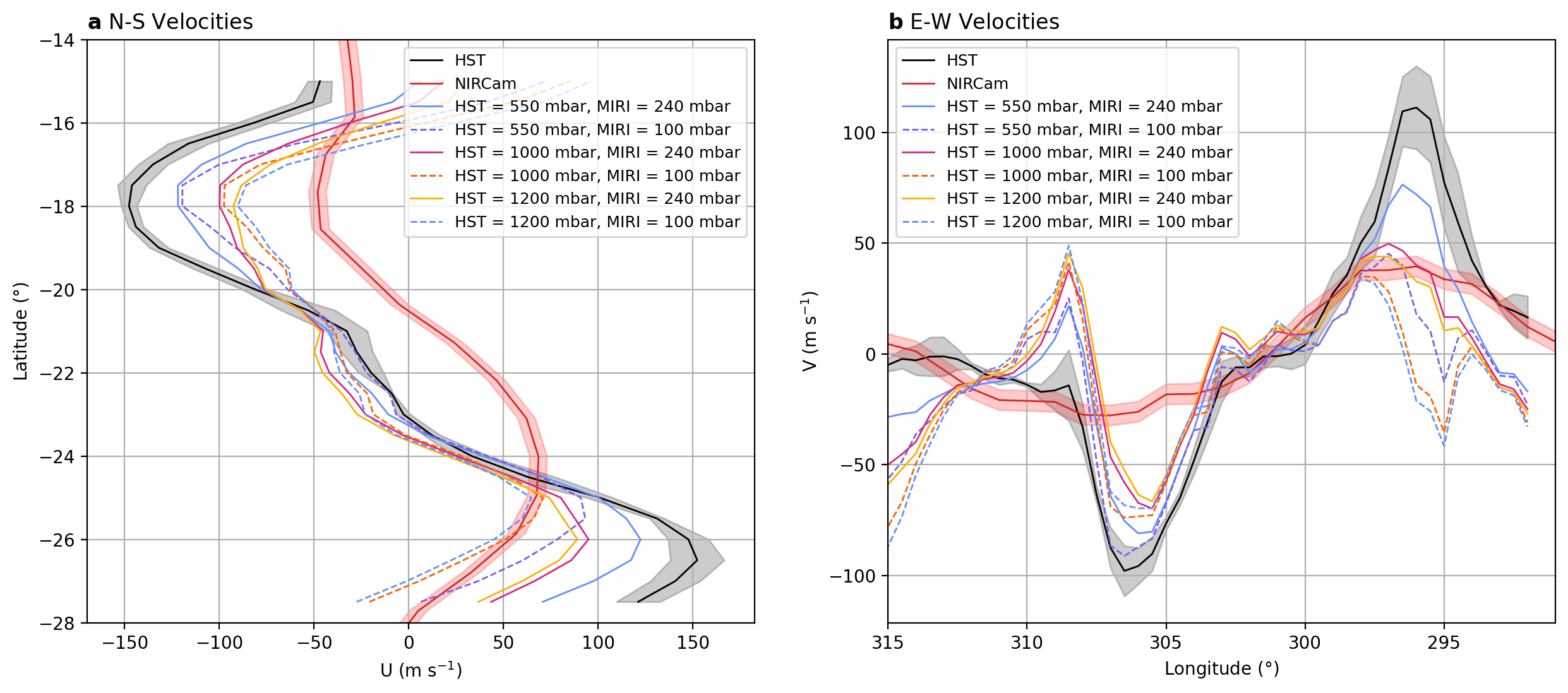}
                \caption{(a) Comparison of the north-south winds passing through the centre of the GRS in August at 301.0$\degree$W and their respective uncertainties inferred from Hubble, NIRCam and the thermal winds derived from the MIRI August data. These MIRI thermal winds are either plotted for 240 mbar or 100 mbar. (b) Similar parameters for the east-west slit centred on 22.5$\degree$S. The Hubble and NIRCam data have been shifted west by 5.5$\degree$ to be centred on the same longitude as the GRS in the MIRI data. Note the considerable drop in velocities between Hubble data (nominally sensitive to 550 mbar) and NIRCam data (nominally sensitive to 240 mbar). Such a large decline in velocity is only observed in the calculated MIRI thermal winds in both the north-south and east-west directions if the Hubble winds are assumed to be located at an altitude of 1200 mbar and the 100 mbar MIRI thermal winds are compared to the NIRCam winds. Note that although uncertainties are omitted from the MIRI-derived winds for clarity, the extent of the uncertainty is represented by the differences between the curves.}
            \label{nircam_comparison}
        \end{figure}
        
        Fig. \ref{nircam_comparison} shows this Hubble and NIRCam wind data. NIRCam displays considerably lower wind velocities compared to Hubble, concurring with the indication that the winds de-spin with altitude. The MIRI inferred thermal winds are also displayed at either 240 or 100 mbar using different assumptions for the prior Hubble wind altitude to determine if any combinations can produce a MIRI thermal wind that is consistent with the measured NIRCam winds. A nominal altitude of 550 mbar for the Hubble data does not permit sufficient loss of velocity with altitude in the MIRI thermal winds to match the NIRCam velocities. In fact, we need to move the Hubble winds to 1200 mbar and the NIRCam winds to 100 mbar for there to be sufficient vertical distance to allow the MIRI thermal winds to decay to match the near-infrared winds in both the north-south and east-west directions. Potentially degeneracies between the temperature profile and the aerosols within the GRS altered the retrieved temperature profile, affecting the retrieved thermal winds. However, this could also be caused by both the Hubble/F631N and NIRCam/F212N filters being sensitive to different altitudes within the GRS than previously thought. Results from past visible and near-infrared retrievals suggest that the main jovian cloud layer may be located deeper than 1,000 mbar \cite{nnnnn, ooooo, ppppp, qqqqq, dahl_2021grscolour_paper}. Since the visible wavelength range is dominated by sunlight scattering off aerosol layers, it is possible that this spectral range could contain contribution from deeper than the 1,000 mbar level. 
        \par
        The horizontal temperature gradients in Equations \ref{wind1} and \ref{wind2} ($\partial T$) were determined by taking the difference in temperature between adjacent spaxels in Fig. \ref{2D_tempos}, $T_0$ and $T_1$ in this case. The uncertainty in this gradient ($\Delta \partial T$) was then a function of the (different) uncertainties in these 2 adjacent points ($\Delta T_0$ and $\Delta T_1$). This was given by:

         \begin{equation}
            \label{delta_grad_equation}
             \Delta \partial T = \left( \Delta T_{0}^{2} + \Delta T_{1}^{2} \right)^{\frac{1}{2}}
         \end{equation}
         
         Error propagation was then used to determine the resulting uncertainty in wind shear due to the uncertainties in both $T$ and $\Delta T$. This ranged from 0.2 m s$^{-1}$ km$^{-1}$ in the troposphere to above 2.0 m s$^{-1}$ km$^{-1}$ in the stratosphere. The thermal wind velocities also had uncertainty originating from the Hubble prior wind fields. Since a median of 3 epochs of data was used to construct this prior, the standard deviation between these 3 datasets was used as the prior uncertainty, typically on the order of 8 m s$^{-1}$. Again, error propagation was used to determine the uncertainties in wind velocities, with temperature, temperature gradient and the prior wind all possessing measurement uncertainties. The resulting velocity uncertainties were smallest at the chosen boundary condition altitude and grew rapidly above and below this, with the uncertainties being of order 10 - 20 m s$^{-1}$ at the tropopause. The uncertainties for both the wind shear and wind velocities can be seen in Supplementary Figure S12. Thermal wind analysis results in significant uncertainties and so should only be used as a guideline for general atmospheric trends.

        \begin{figure}
            \centering
            \includegraphics[width=1.0\textwidth]{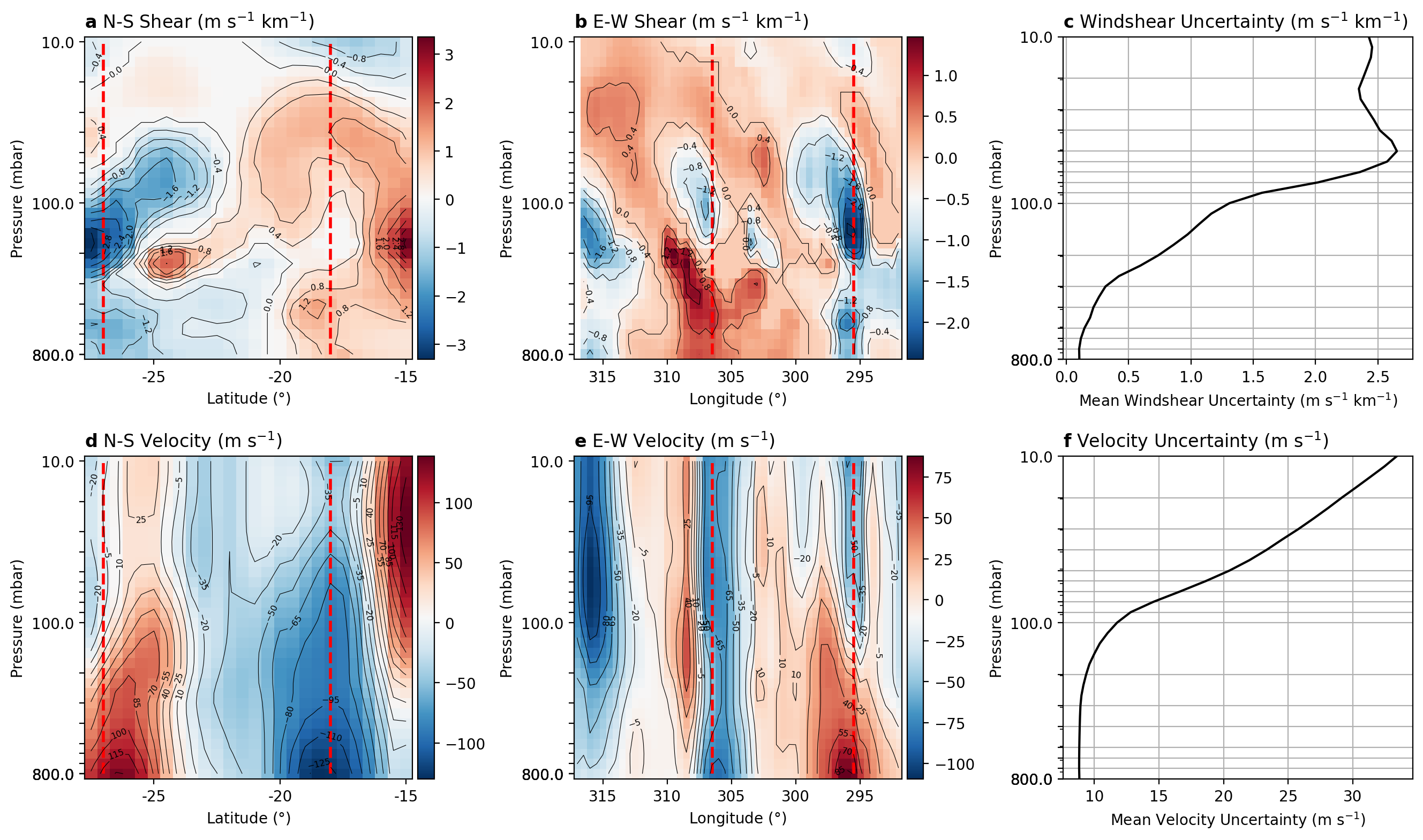}
            \caption{(a) Thermal wind shear in the north-south direction for the August epoch, centred on 22.5$\degree$S and inferred from the retrieved temperatures in Fig. \ref{2D_tempos}a and Equation \ref{wind1}. (b) Thermal windshear in the east-west direction centred on 301.0$\degree$W inferred from the retrieved temperatures in Fig. \ref{2D_tempos}b and Equation. \ref{wind2}. (c) Mean of the uncertainties in windshear plotted as a function of altitude. 2D maps of this uncertainty can be seen in Supplementary Figure S12. An initially well-constrained uncertainty grows rapidly around the 100 mbar tropopause. (d) Inferred thermal wind structure in the north-south direction by integrating Equation \ref{wind1} with height and assuming an altitude of 1,200 mbar for the HST wind measurements. (e) Inferred east-west thermal wind velocities determined by integrating Equation \ref{wind2} with height. (f) Mean of the velocity uncertainties with respect to altitude, note the uncertainties grow rapidly with increasing altitude from the 1,200 mbar boundary condition. The red dashed lines denote the boundaries of the GRS vortex in each case. For the windshear and thermal wind plots, white corresponds to 0 m s$^{-1}$ km$^{-1}$ and 0 m s$^{-1}$ respectively.}
            \label{july_thermal_windshear}
        \end{figure}

        The derived windshears can be seen in Fig. \ref{july_thermal_windshear}a and b. Considerable differences are observed for the windshears in the north-south direction. Around the 100-mbar level, the southern edge of the vortex at 25$\degree$S in Fig. \ref{july_thermal_windshear}a displayed a wind-shear that was almost twice that of the corresponding northern region. Below this and within the collar outside the dashed vertical lines denoting the boundaries of the GRS peak wind velocities (determined using Hubble) the magnitude of the northern and southern windshear were comparable. Like the temperature, NH$_3$, PH$_3$ and aerosol north-south asymmetries described in Section \ref{asymmetries}, these windshear measurements are concurrent with the proposed tilt of the GRS. \citeA{sada_1996grsvoyager_paper} suggested the southern portion of the GRS could flow deeper than the north to explain observed asymmetries in Voyager 1 and 2 data. The implied deeper flow could be responsible for the asymmetries seen in the zonal wind shear in Fig. \ref{july_thermal_windshear}, with the higher-altitude northern limb decaying over a broader range altitudes than in the south.
        \par
        Fig. \ref{july_thermal_windshear} c and d displays 2D plots of the inferred north-south and east-west thermal winds, assuming a cloud-top pressure boundary condition of 1,200 mbar. Velocities decay with altitude as expected. However, dissipation of the cold temperature anomaly in the retrieved temperature structure at shallower altitudes than the 200 mbar pressure level prevents the GRS wind velocities from dissipating at the tropopause (100 mbar), as the dynamical nature of the vortex would imply \cite{60}. It is possible that the vigorous vortex dynamics may allow the GRS to overshoot the tropopause into the stratosphere \cite{sanchez-lavega_2018grsjunocam_paper, anguiano-arteaga_2021grsaerosol_paper}. Indeed the windshear plots indicate the persistence of strong wind shear around the 100 mbar level. However, it is highly unlikely that the GRS winds persist up to 10 mbar. The observed peak horizontal temperature gradients at 400 mbar can be seen in Table \ref{required_thermal_winds} and are compared to the temperature gradient required to force the thermal wind to reach 0 m s$^{-1}$ at 100 mbar, assuming a depth of 1200 mbar for the Hubble winds. A horizontal temperature gradient nearly 2.0 times greater is required to satisfy this constraint within all of the GRS cardinal boundaries. Assuming a size of 14,000 $\times$ 11,000 km for the GRS, these gradients would correspond to a cold-temperature anomaly of -9 to -14 K within the GRS compared to the surroundings. Such values are observed in a localised region above the 200 mbar pressure level. However, below this pressure level the magnitude of the temperature anomaly fluctuates considerably. Such large temperature gradients have not been observed at the required broad range of altitudes and positions. This could be due to the limited spatial resolution of the MIRI data, the variable information content from nadir spectroscopy at the required depths, the temperature, NH$_3$, PH$_3$ and aerosol degeneracies described in Section \ref{degeneracy} or the GRS thermal anomaly and resulting windshear persisting into the lower-stratosphere. The latter would be consistent with the high-altitude aerosols implied by the near-infrared observations seen in Fig. \ref{general_context_figures}a and b.
            
        \begin{table}
            \centering
            \begin{tabular}{c c c c c}
                \hline
                Position & $V$ (m s$^{-1}$) & $\phi_g$ ($\degree$S) & $R$ (mK km$^{-1}$) & $M$ (mK km$^{-1}$) \\
                \hline
                North & -147.60 ± 4.17 & 18.00 & 1.72 ± 0.05 & 0.85 ± 0.14\\
                South & 152.26 ± 14.23 & 26.50 & -2.57 ± 0.24 & -1.44 ± 0.15\\
                West & -97.97 ± 11.41 & 22.50 & -1.42 ± 0.17 & -0.79 ± 0.14\\
                East & 111.17 ± 18.77 & 22.50 & 1.61 ± 0.27 & 1.04 ± 0.15\\
                \hline
            \end{tabular}
            \caption{Maximum east-west and north-south velocities at each of the four cardinal boundaries of the vortex ($V$) measured within a cross section of the median HST wind data. Temperature gradients are inferred ($R$) from these by integrating equations \ref{wind1} and \ref{wind2} with respect to height and assuming the wind speeds reach 0 m s$^{-1}$ at the tropopause. These are compared to the maximum thermal gradients measured ($M$) at 400 mbar.}
            \label{required_thermal_winds}
        \end{table}

        Analysis of the vertical atmospheric stability could be conducted using the Brunt-Väisälä Frequency ($N^{2}$):
            
        \begin{equation}
            N^2 = \frac{g}{T} \left(\frac{\partial T}{\partial z} + \Gamma_d\right)
        \end{equation}

         where $\frac{\partial T}{\partial z}$ is the vertical temperature gradient and $\Gamma_d$ is the Dry Adiabatic Lapse Rate (DALR), given by $\Gamma_d = \frac{g}{c_p}$, where $c_p$ is the atmospheric heat capacity, assumed to be 13.07 J K$^{-1}$ g$^{-1}$ using the bulk composition introduced in Section \ref{NEMESIS}. Statically unstable regions of the atmosphere have $N^2 < 0$ while statically stable regions have $N^2 > 0$. In the latter case, $N$ represents the frequency of buoyancy oscillations. The uncertainty in $N^2$ was determined by error propagation using the uncertainties in temperature and the temperature gradient, given by Equation \ref{delta_grad_equation}. A map of this uncertainty in $N^2$ can be seen in Supplementary Figure S13.

        \begin{figure}
            \centering
            \includegraphics[width=1.0\textwidth]{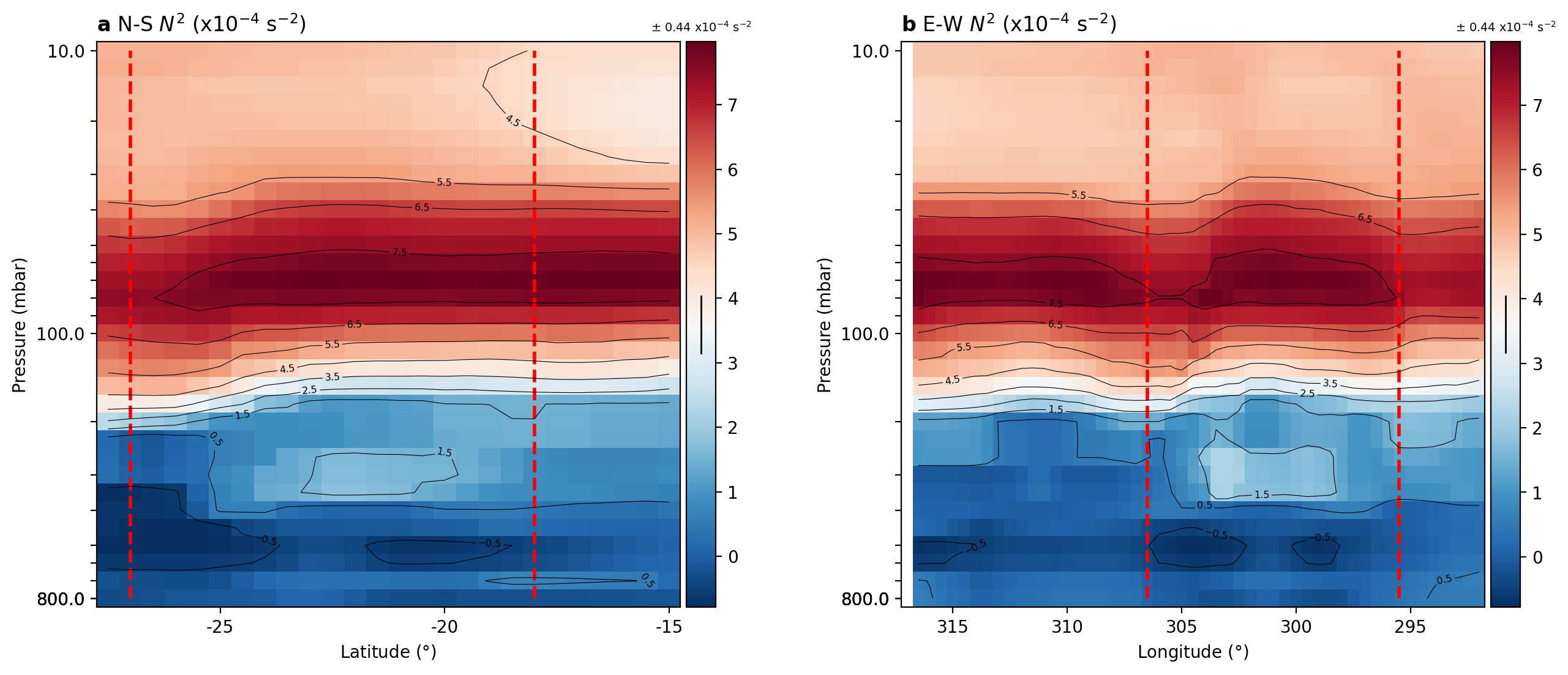}
            \caption{Buoyancy frequency inferred from the August MIRI data for (a) the north-south slit and (b) the east-west slit. The vertical dashed lines indicate the boundaries of the GRS vortex. The median of the 1$\sigma$ propagated buoyancy frequency uncertainty in this altitude range is visible as a vertical black line within each of the colourbars and the numerical value of these uncertainties is given above the colour bar.}
            \label{thermal_stability_july}
        \end{figure}
        
         Fig. \ref{thermal_stability_july} displays the inferred $N^2$ maps. The median of the uncertainties is displayed in the colour bar of each plot. At this small range of pressures, little variation in $N^{2}$ would be expected to be seen. The GRS was observed to be indistinguishable compared to the surroundings in the east-west plot, with regions of negative $N^2$ being observed equally inside and outside the GRS. In the north-south plot, a region of elevated $N^{2}$ was observed on the southern limb of the ring of peak GRS velocities. This coincided with the region of presumed subsidence south of the GRS, responsible for the observed arc of warmer temperatures and depleted NH$_3$. The lack of a large-scale convective instability across the GRS has important implications for both the cloud forming processes inside the vortex as well as the distribution of PH$_{3}$ across it, which was shown in Section \ref{phosphine} to be in considerable excess above the GRS compared to the surroundings. Potentially, the lack of any vigorous convective activity within the vortex may allow this molecule to accumulate, while the excess aerosol opacity shields the molecule from the UV light that would normally photolyse and remove it as discussed in Section \ref{asymmetries}. It is difficult to envisage the thick cloud banks of the GRS being generated through convective processes in such a convectively stable region of the atmosphere. Potentially, the GRS cold-temperature anomaly could force NH$_3$ vapour to condense into stratiform cloud layers within the vortex as outlined in Section \ref{ammonia}.

    \subsection{NH$_3$ ice non-detection}
        \label{nh3_non_de}

        \begin{figure}
            \centering
            \includegraphics[width=\textwidth]{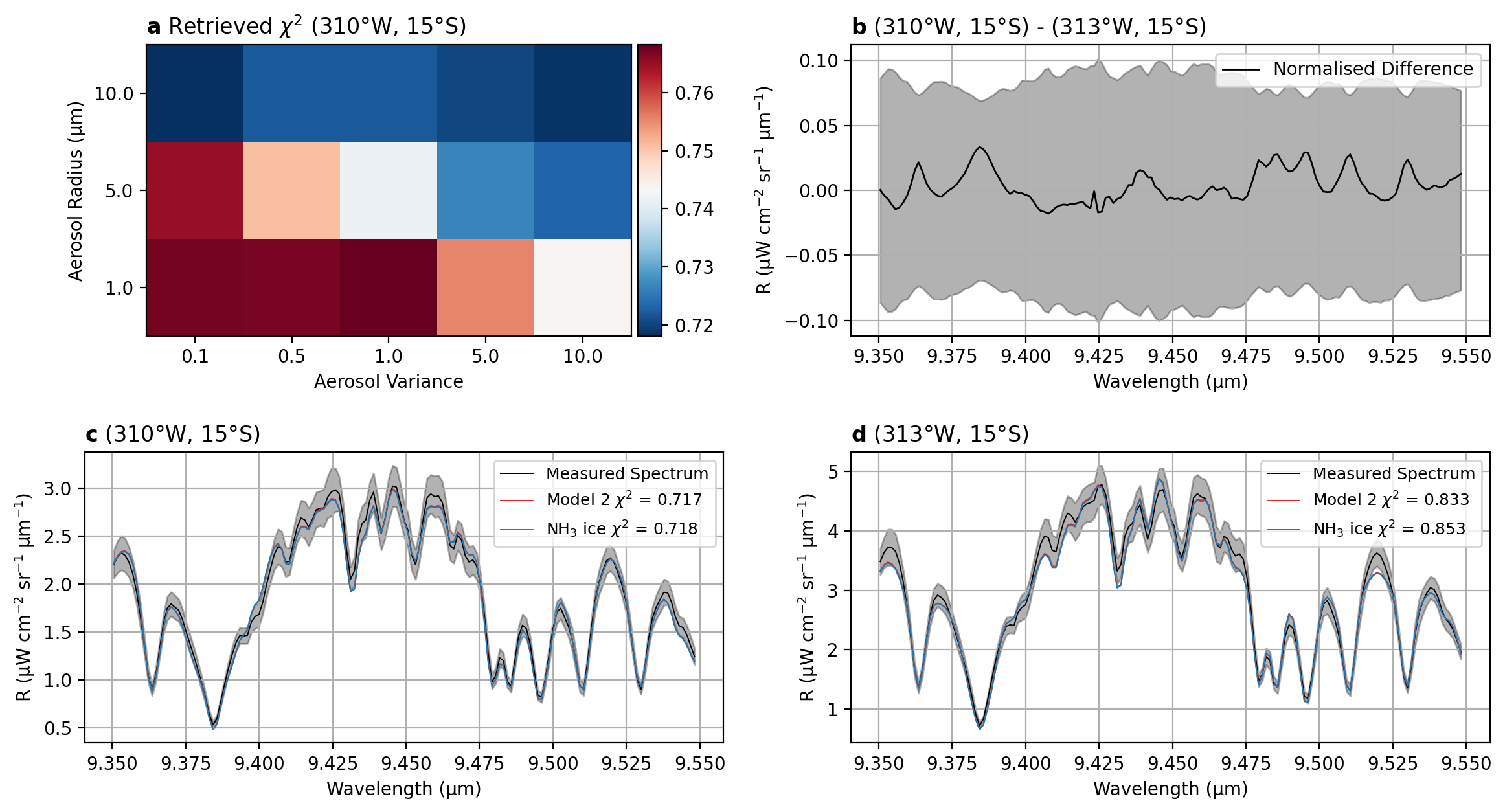}
            \caption{Results for the modelling of the 9.46 µm NH$_3$ ice feature, both inside the VICI 1 storm (c) and outside (d). (a) Retrieved $\chi^2$ inside VICI 1. A variable refractive index based on the properties of NH$_3$ ice was assumed. A variety of Radii ($R$) and Variance ($V$) combinations were examined for this model and $R$ = 10.0 µm, $V$ = 0.1 provided the best goodness-of-fit. (b) Difference between normalised spectra for inside VICI 1 (310$\degree$W, 15$\degree$S) and outside (313$\degree$W, 15$\degree$S). The extent of the NH$_3$ ice feature varies with particle radius, with larger radii potentially generating a feature that fills the entire wavelength range shown here. The mean uncertainty for these two positions is plotted as a shaded region. Little variation was observed between these two spectra. (c) Within VICI 1, the $R$ = 10.0 µm, $V$ = 0.1 NH$_3$ ice model was then compared to model 2, previously used for the retrieval maps presented in Section \ref{results} and containing no NH$_3$ ice. Constant real and imaginary refractive indices of 1.2 and 0.005 respectively were assumed for model 2. Little variation can be seen between these two models, indicative that NH$_3$ ice is not required to fit the measured spectrum here. (d) Similar comparison of these two models outside VICI 1. Here, the constant refractive index model consistently provided a better fit to the measured spectrum than the NH$_3$ ice model across the entire wavelength range.}
            \label{nh3_ice_nondetection}
        \end{figure}

        ECCM models of the jovian atmosphere \cite{irwin_book_latestedition} predict the presence of an NH$_3$ ice cloud centred around 800 mbar \cite{eeeee, weidenschilling_1973jupcloud_paper}. Pure NH$_3$ ice produces two distinctive absorption bands; the $\nu_3$ band at 2.96 µm and the $\nu_2$ band at 9.46 µm \cite{wong_nh3icev2_paper}. Detection of these bands have been elusive and have only been observed across 1\% of the jovian surface \cite{baines_nh3icev3_paper}. Neither of these features are detectable from the ground \cite{wong_nh3icev2_paper} and Voyager/IRIS \cite{jjjjj, hhhhh, iiiii} failed to detect the $\nu_2$ feature in both the Equatorial Zone and the North Tropical Zone, despite having sufficient spectral resolution and sensitivity to do so. The $\nu_3$ feature was first detected in disc-averaged ISO data \cite{ggggg, zzzz}. This was then subsequently mapped within the GRS wake using Galileo/NIMS \cite{baines_nh3icev3_paper}. Zonal mapping of the $\nu_2$ feature across the jovian disc was carried out using Cassini/CIRS \cite{wong_nh3icev2_paper}. New Horizons detected a further NH$_3$ ice absorption feature at 1.99 µm within the GRS wake \cite{lllll}. NH$_3$ ice is believed to have a lifetime on the order of days, before either being destroyed or coated by other species that obscure the spectral signature of this species \cite{baines_nh3icev3_paper}. Therefore, the only areas where NH$_3$ ice detection would be expected are within regions of vigorous upwelling, where the uplift of NH$_3$ ice to altitudes where it can be detected outstrips the mechanisms that obscure it.
        \par
        The spectrum from within the VICI 1 storm was searched for traces of the $\nu_2$ band at 9.46 µm. This region was chosen due to its high-albedo in the visible amateur images and the 890 nm image shown in Fig. \ref{general_context_figures}b, suggesting the presence of fresh clouds elevated in altitude compared to the surroundings and containing less colouring agent. The PH$_3$ excess within this instability further supports the theory that this is a region of convective uplift. Two spectra were modelled, one at (310$\degree$W, 15$\degree$S) within VICI 1 and one displaced 3 degrees west to probe the wider GRS wake. The constant refractive index model of n = 1.2 + i0.005 (model 2) used for the retrieval maps of section \ref{results} was compared to a model possessing the optical properties of NH$_3$ ice. The NH$_3$ ice feature has a strong dependence on the particle size, with smaller radii generating a more sharply peaked absorption feature and larger radii generating a broader feature \cite{wong_nh3icev2_paper}. This radius is loosely constrained however, with the $\nu_3$ band generally being best fitted by 10 µm particles \cite{baines_nh3icev3_paper} and the $\nu_2$ band being best fitted with 1 µm particles \cite{wong_nh3icev2_paper}. This disparity resulted in us testing 15 NH$_3$ ice models with radii and variances ranging from 1.0 - 10.0 µm and 0.1 - 10.0 respectively. This can be seen in Fig. \ref{nh3_ice_nondetection}a. The model containing particles of $R$ = 10.0 µm, $V$ = 0.1 provided the best fit to the measured spectrum. The 9.35 - 9.55 µm spectrum of this model was then compared to model 2 and the measured spectrum within VICI 1 (Fig. \ref{nh3_ice_nondetection}c). Little difference can be seen between the two models inside VICI 1. Outside (Fig. \ref{nh3_ice_nondetection}d), the NH$_3$ ice model displays a significantly poorer fit compared to model 2. Finally, the two spectra were normalised and the difference between these two spectra can be seen in Fig. \ref{nh3_ice_nondetection}b. There was no indication of a residual resembling the broad NH$_3$ ice absorption feature, with all fluctuations being within one standard deviation. Overall, Fig. \ref{nh3_ice_nondetection} indicates that modelling NH$_3$ ice is not required to reproduce the measured spectrum. If NH$_3$ ice was present in this data, it was either obscured by other species or was too subtle to be detected.

\section{Conclusions}
\label{conclusions}

    This initial survey of the atmosphere surrounding the GRS has shown the potential of JWST/MIRI to characterise this dynamic region of Jupiter's atmosphere over 18 days in July and August 2022. Observing one of the brightest targets to date with MIRI required the development of new, custom-made calibration algorithms and techniques to allow analysis at a range of wavelengths. The exposure time, using 4 groups was shorter than the 5 recommended by STScI. We however encountered no problems calibrating this data with the calibration pipeline, paving the way for future observations of Jupiter and other comparably bright targets. The well-mapped forest of molecular lines in the spectrum also provided an opportunity to assess the wavelength calibration of the instrument. This led to the discovery of lingering offsets between adjacent slices in the MRS FOV that were corrected using data from Jupiter and Saturn (GTO 1247), deriving the FLT-5 wavelength calibration solution that has been implemented in the calibration pipeline as standard in calibration reference contexts newer than \verb|jwst_1112.pmap| \cite{argyriou_wavecal_paper}.
    \par
    A series of flat-fields for each band and tile also had to be developed for these observations due to lingering artefacts remaining in the post-pipeline data. Interestingly Saturn \cite{fletcher_jwstsaturn_paper} required a different flat-field and no such artefacts were observed in the darker background frames taken alongside the July MIRI observations. It is possible that the JWST flat-field is dependent on the brightness of the target illuminating the detector. Excessive fringing and significant saturation issues beyond 10.75 µm limited us to wavelengths below this threshold. Despite this, calibration of the MIRI wavelengths and removal of the artefacts revealed a stunning array of atmospheric features in the 7.30 - 10.75 µm range as summarised below:

    \begin{enumerate}
        \item \textbf{Inhomogeneous temperature structure in the Stratosphere}\\
        A zonal structure of vertically alternating warm and cold temperature bands were observed to be centred on 17$\degree$S at altitudes of 30 mbar, 8 mbar and 3 mbar respectively. This most likely corresponded to the off-equatorial temperature anomalies associated with the jovian Quasi-Quadrennial Oscillation (QQO). In particular, the observation of a warm temperature anomaly north of the MIRI FOV at 30 mbar which inverted to be a cold temperature anomaly at 8 mbar and further reverted to a warm anomaly at 3 mbar concur with the observations made by \citeA{56} 4 years prior to the MIRI observations. The MIRI observations of this feature were consistent with the 4-year period of the QQO.
        \par
        The presence of two hot-spots at 3 mbar that co-move with the GRS below and potentially more hot-spots at similar latitudes across the jovian disc seen by VLT/VISIR may indicate a coupling between the turbulence of the GRS in the troposphere to the stratosphere above. This coupling would have the potential to generate a series of atmospheric waves, the presence of which could be responsible for the generation of these hot-spots. However, further modelling of these wave phenomena is necessary before more speculation can be made as to the nature of these features.

        \item \textbf{Thermal wind structure of the GRS}\\
        The ability of JWST/MIRI spectroscopy to sense every altitude in the 800 - 100 mbar range in the troposphere allowed the thermal wind equations to be used to infer the dynamical structure of the GRS. This windshear displayed a north-south asymmetry that concurs with the suggestion from observations of gaseous species of enhanced upwelling in the north of the vortex and subsidence in the south. Calculated thermal winds using prior velocities derived by Hubble/WFC3's F631N filter were also consistent with the velocities determined by JWST/NIRCam's F212N filter. However, consistency was only achieved by assuming that the Hubble and NIRCam winds were located at 1,200 mbar and 100 mbar respectively, a larger altitude separation than is typically assumed for these instrument/filter combinations. Although the spatial resolution of JWST/MIRI is lower than some ground-based observatories, which may prevent the largest horizontal temperature gradients from being observed, there is evidence that visible wavelengths (such as those probed by Hubble/WFC3) could be probing altitudes deeper than the 1,000 mbar pressure level \cite{nnnnn, ooooo, ppppp, qqqqq, dahl_2021grscolour_paper}. The GRS is also a highly anomalous region of Jupiter’s atmosphere and contribution functions that apply to the wider jovian atmosphere may not apply to the GRS. Ultimately the inability of the thermal wind equation to constrain the inferred primary circulation speeds to rest at the tropopause is likely to be a combination of these two factors. The derived buoyancy frequency within the GRS is indistinguishable to the surroundings, despite expectations that the upper part of the anticyclone would be stable to convection.

        \item \textbf{Inhomogeneous temperature/compositional structure in the GRS}\\
        North-south asymmetries were observed in the temperature,  NH$_3$, PH$_3$ and aerosol structure. This concurs with previous suggestions that the GRS may tilt in the north-south direction, with possible upwelling in the northern collar and subsidence in the southern collar. It is possible that the GRS flows deeper in the south of the vortex \cite{sada_1996grsvoyager_paper}, which would account for the north-south asymmetry observed in the measured zonal thermal windshear, with the windshear in the south being almost twice that in the north. In addition to this, small-scale variability was observed within the cold-temperature anomaly of the GRS which was observed to change over the 18 days separating observations. Unfortunately, due to degeneracies inherent in the 7.30 - 10.75 µm region between the retrieved parameters, it is difficult to determine if these effects are the result of temperature or NH$_3$ variability. One example is the small hot-spot located at 24$\degree$S, which persisted at a broad range of tropospheric altitudes (600 – 300 mbar) and resembles the coherent warm core observed by \citeA{fletcher_2010grs_paper}. Potentially this could be the same feature or it could be a small internal vortex barely resolved by the visual observations. The anti-correlation between NH$_3$ and aerosol opacity could be due to any excess NH$_3$ condensing into the thick banks of clouds known to dominate the GRS vortex.
        \item \textbf{Temperature, aerosol and gaseous degeneracies}\\
        Although temperature and composition are well constrained in the upper troposphere (p $<$ 800 mbar), we have been unsuccessful in breaking the degeneracies inherent between the retrieved parameters within the 7.30 - 10.75 µm spectral region for altitudes deeper than the 800 mbar pressure level. Juno/MWR suffers a similar issue with degeneracies between deep temperature and NH$_3$ VMR. Potentially, by combining the results from JWST/MIRI and Juno/MWR, it could be possible to disentangle the contributions of these two species. Ultimately higher spatial and spectral resolution may be required in the far-infrared range of the jovian spectrum (15 - 1,000 µm). This would allow the tropospheric temperature structure of this region to be determined by fitting the H$_2$-He continuum without the complicating influence of NH$_3$ absorption. Such observations cannot be made from the ground due to telluric absorption.

        \item \textbf{The effect of aerosols on the PH$_3$ distribution within the GRS vortex}\\
        Unlike NH$_3$, the PH$_3$ VMR was observed to be in considerable excess over the GRS compared to the surroundings. This excess was also determined to be coincident with the region of red chromophore material inside the ring of peak wind velocities surrounding the vortex. Qualitatively, the retrieved aerosol opacity displays excess in the same locations above the GRS as PH$_3$ does. It is possible that the elevated aerosol opacity within the GRS could shield the PH$_3$ from the UV light that would normally photolyse and remove this molecule. The stability of this region against convection could allow sufficient time for this molecule to accumulate above the GRS, with the anticyclonic nature of the vortex potentially acting to further concentrate this molecule. The VICI 1 storm most likely has not existed for sufficient time to develop a similar high-altitude (above 500 mbar) PH$_3$ feature.

        \item \textbf{Characteristics of VICI 1}\\
        The fortunate presence of a moist convective plume (VICI 1) in the August epoch, within the already unstable GRS wake provides a unique opportunity to probe the deeper composition of Jupiter’s atmosphere. Despite the non-detection of NH$_3$ ice within this region, the inferred excess of PH$_3$ is suggestive of a vigorously convective cell that could have driven other molecules normally found below the weather layer to an altitude where we are capable of detecting them.

        \item \textbf{Non-detection of NH$_3$ ice}\\
        The non-detection of NH$_3$ ice in the GRS wake in this study may indicate that VICI 1 is not powerful enough to have lofted fresh NH$_3$ condensate to the upper troposphere. Ultimately, spatial and temporal variation of this species is still poorly understood. It is possible that the NH$_3$ ice particles may be rapidly coated by a currently unidentified species that removes the expected absorption features over the course of a few days \cite{baines_nh3icev3_paper}. A similar mechanism may also account for the rare detection of NH$_3$ ice in Saturn's atmosphere too \cite{kkkkk}. Further modelling of the chemical interaction of NH$_3$ ice molecules with the surrounding atmosphere is necessary to better understand the regular absence of NH$_3$ ice signatures from the jovian spectrum. Additional IFU observations by both JWST/NIRSpec and JWST/MIRI would also enable analysis of both the spatial and temporal variability of NH$_3$ ice, yielding vital information regarding Jupiter's vertical dynamics and aerosol chemistry.

    \end{enumerate}
    The algorithms required to process MIRI/MRS data are still evolving. In future, it may be possible to mitigate the fringing beyond 10.75 µm. This would allow mapping of the C$_2$H$_2$ and C$_2$H$_6$ emission features in channel 3 as well as enabling a search for minor species such as C$_6$H$_6$, HCN and CO$_2$. A future campaign jointly fitting both the MIRI observations alongside Juno/MWR GRS observations acquired in 2017 \cite{bolton_2021mwr_paper} would potentially break the degeneracies inherent in both datasets, allowing a greater understanding of the vertical structure of this vortex as it continues to evolve.
    \par
    The GRS is a highly dynamic region of Jupiter's atmosphere that would benefit from regular JWST observations. Mechanisms such as SEB fades, the GRS's interaction with other vortices such as oval BA and the shrinkage \cite{7} of the GRS would be particularly interesting to study with this new observatory. Annual observations made of the GRS using Hubble since 2015 as part of the OPAL and WFCJ programmes have proved to be invaluable in tracking dynamical changes at a single altitude within this vortex over time. A similar campaign of long-term observations with JWST could revolutionise our understanding of jovian vortices and atmospheric dynamics in general.

\section{Open Research}

    Level-3 calibrated GRS MIRI/MRS data (GTO 1246) from the standard pipeline are available directly from the MAST archive \cite{MAST1246_MIRI}. The Hubble observations used for comparison were acquired by programme: GO-16913 \cite{MAST_HST_GRS}.
    \par
    The position of the GRS was estimated based on amateur observer submissions to PVOL \cite{PVOL_data}. The visual context observations taken on 2022-08-15 were made by Isao Miyazaki. An archive of amateur images of Jupiter including the one used in this study can be found online \cite{ALPO_data}.
    \par
    The PICV3 software used to derive GRS wind velocities from the NIRCam data is available to download \cite{PICV3_software}. The ACCIV software used to derive GRS wind velocities from the HST data is also available to download \cite{ACCIV_software}.
    \par
    The NEMESIS suite of radiative transfer and spectral inversion software \cite{irwin_nemesis_paper} is open-access and available for download \cite{NEMESIS_software}. The JWST calibration pipeline is available as a Python module \cite{JWSTpipeline_software}. Version 1.11.4. was used for this study. The bespoke pipeline and data processing code developed during this study are available to download \cite{JWSTbespoke_software}.
    \par
    The PlanetMapper software used to assign longitude and latitude grids to the VLT/VISIR and ground-based amateur data can be found online \cite{king_2023planetmapper_paper}.
    \par
    The data products produced in this study and the code used to generate these products are available online \cite{2024harkett_data}.

\acknowledgments

We would like to thank David Law of the JWST support team for his incredible work generating new wavelength-calibration solutions as well as his patience in answering numerous questions. Additionally, we are grateful to the GTO and ERS 1373 teams for their expertise and numerous helpful discussions. We also appreciate the detailed and high-quality feedback from two anonymous reviewers.
\par
Jake Harkett was supported by an STFC studentship. Leigh N. Fletcher, Oliver R.T. King and Michael T. Roman were supported by a European Research Council Consolidator Grant (under the European Union’s Horizon 2020 research and innovation programme, grant agreement No. 723890). Henrik Melin was supported by the STFC JWST Fellowship (ST/W001527/1). Pablo Rodríguez-Ovalle was supported by a Université Paris-Cité contract. Thierry Fouchet was supported by the Grant ANR-21-CE49-0020-01. Ricardo Hueso and Agustín Sánchez-Lavega were supported by grant PID2019-109467GB-I00 funded by MCIN/AEI/10.13039/501100011033/ and were also supported by Grupos Gobierno Vasco IT1742-22. Part of this research was carried out at the Jet Propulsion Laboratory, California Institute of Technology, under a contract with the National Aeronautics and Space Administration (80NM0018D0004).
\par
This work is based on observations made with the NASA/ESA/CSA James Webb Space Telescope. The data were obtained from the Mikulski Archive for Space Telescopes at the Space Telescope Science Institute, which is operated by the Association of Universities for Research in Astronomy, Inc., under NASA contract NAS 5-03127 for JWST. These observations are associated with programme 1246, led by PI; Leigh N. Fletcher and programme 1373, led by co-PIs; Imke de Pater and Thierry Fouchet. Hubble observations were made as part of programme: GO16913 (PI: Michael H. Wong) and the amateur ground-based observations were made by Isao Miyazaki. VLT observations were collected at the European Southern Observatory under ESO programme 108.223F.001. For the post-processing of the MIRI data, this research used the ALICE High Performance Computing Facility at the University of Leicester. For the purpose of open access, the author has applied a Creative Commons Attribution (CC BY) licence to the Author Accepted Manuscript version arising from this submission.

\bibliography{references}

\end{document}